\documentclass[reqno,11pt]{amsart}

\usepackage{amssymb,graphics,graphicx,wrapfig}

\newcommand{\floor}[1]{\left\lfloor #1 \right\rfloor}
\def\eps{\varepsilon}

\def\N{\bbN}

\def\be{\begin{equation}}
\def\ee{\end{equation}}
\def\ba{\begin{align}}
\def\bm{\begin{multline}}
\def\bfig{\begin{figure}[htb]}
\def\efig{\end{figure}}

\numberwithin{equation}{section}
\newtheorem{theorem}{Theorem}[section]
\newtheorem{proposition}[theorem]{Proposition}
\newtheorem{lemma}[theorem]{Lemma}
\newtheorem{corollary}[theorem]{Corollary}

\DeclareMathSymbol{\leqslant}{\mathalpha}{AMSa}{"36}
\DeclareMathSymbol{\geqslant}{\mathalpha}{AMSa}{"3E}
\DeclareMathSymbol{\doteqdot}{\mathalpha}{AMSa}{"2B}
\DeclareMathSymbol{\circlearrowright}{\mathalpha}{AMSa}{"08}
\DeclareMathSymbol{\subsetneq}{\mathalpha}{AMSb}{"28}
\DeclareMathSymbol{\supsetneq}{\mathalpha}{AMSb}{"29}
\renewcommand{\leq}{\;\leqslant\;}
\renewcommand{\geq}{\;\geqslant\;}

\newcommand{\dd}{{\rm d}}
\newcommand{\e}[1]{\,{\rm e}^{#1}\,}
\newcommand{\ii}{{\rm i}}
\newcommand{\sumtwo}[2]{\sum_{\substack{#1 \\ #2}}}

\DeclareMathOperator*{\Union}{\text{\Large$\cup$}}
\newcommand{\Uniontwo}[2]{\Union_{\substack{#1 \\ #2}}}
\DeclareMathOperator*{\inter}{\text{\large$\cap$}}
\DeclareMathOperator*{\Inter}{\text{\Large$\cap$}}

\def\Tr{{\operatorname{Tr\,}}}

\newcommand{\compl}{{\text{\rm c}}}

\newcommand{\upchi}{\raise 2pt \hbox{$\chi$}}

\def\writefig#1 #2 #3 {\rlap{\kern #1 truecm \raise #2 truecm
\hbox{#3}}}

\newcommand{\caN}{{\mathcal N}}

\newcommand{\caR}{{\mathcal R}}
\newcommand{\caS}{{\mathcal S}}

\newcommand{\bbN}{{\mathbb N}}

\newcommand{\bbR}{{\mathbb R}}

\newcommand{\bbZ}{{\mathbb Z}}

\newcommand{\bsk}{{\boldsymbol k}}

\newcommand{\bsn}{{\boldsymbol n}}

\newcommand{\bsx}{{\boldsymbol x}}

\newcommand{\bsvarrho}{{\boldsymbol\varrho}}

\begin{document}


\title{Spatial random permutations and infinite cycles}

\author{Volker Betz and Daniel Ueltschi}
\address{Volker Betz and Daniel Ueltschi \hfill\newline
\indent Department of Mathematics \hfill\newline
\indent University of Warwick \hfill\newline
\indent Coventry, CV4 7AL, England \hfill\newline
{\small\rm\indent http://www.maths.warwick.ac.uk/$\sim$betz/} \hfill\newline
{\small\rm\indent http://www.ueltschi.org}
}
\email{v.m.betz@warwick.ac.uk, daniel@ueltschi.org}

\maketitle

\begin{quote}
{\small
{\bf Abstract.}
We consider systems of spatial random permutations, where permutations are weighed according to the point locations.
Infinite cycles are present at high densities.
The critical density is given by an exact expression.
We discuss the relation between the model of spatial permutations and the ideal and interacting quantum Bose gas.
}  

\vspace{1mm}

\vspace{1mm}
\noindent
{\footnotesize {\it 2000 Math.\ Subj.\ Class.:} 60K35, 82B20, 82B26, 82B41.}
\end{quote}

{\footnotesize\tableofcontents}

\section{Introduction}

This article is devoted to random permutations on countable sets that possess a {\it spatial structure}.
Let $\bsx$ be a finite set of points $x_1, \dots, x_N \in \bbR^d$, and let $\caS_N$ be the set of permutations of $N$ elements.
We are interested in probability measures on $\caS_N$ where permutations with long jumps are discouraged.
The main example deals with ``Gaussian" weights, where the probability of $\pi \in \caS_N$ is proportional to
\[
\exp \Bigl\{ -\frac1{4\beta} \sum_{i=1}^N |x_i - x_{\pi(i)}|^2 \Bigr\},
\]
with $\beta$ a parameter.
But we are also interested in more general weights on permutations.
In addition, we want to allow the distribution of points to be random and to depend on the permutations.

We are mostly interested in the existence and properties of such measures in the thermodynamic limit.
That is, assuming that the $N$ points $\bsx$ belong to a cubic box $\Lambda \subset \bbR^d$, we consider the limit $|\Lambda|, N \to \infty$, keeping the density $\rho = N/|\Lambda|$ fixed.
The main question is the possible occurrence of {\it infinite cycles}.
As will be seen, infinite cycles occur when the density is larger than a critical value.

Mathematicians and physicists have devoted many efforts to investigating properties of non-spatial random permutations, when all permutations carry equal weight.
In particular, a special emphasis has been put on the study of longest increasing subsequences \cite{AD, BDJ} and their implications for such diverse areas as random matrices \cite{BDJ, Ok2000}, Gromov-Witten theory \cite{Ok2005} or polynuclear growth \cite{FSP}, and spectacular results have been obtained. 
The situation is very different for random permutations involving spatial structure; 
we are only aware of the works \cite{Kik,KDS,Fic} (and \cite{GRU1}).
This lack of attention is odd since spatial random permutations are natural and appealing notions in probability theory;
it becomes even more astonishing when considering that they play an important r\^ole in the study of quantum bosonic systems:
Feynman \cite{Fey} and Penrose and Onsager \cite{PO} pointed out the importance of long cycles for Bose-Einstein condensation, and later S\"ut\H o clarified the notion of infinite cycles, also showing that infinite, macroscopic cycles are present in the ideal Bose gas \cite{Suto1,Suto2}.
These works, however, never leave the context of quantum mechanics.
We believe that the time is ripe for introducing a general mathematical framework of spatial random permutations.
The goal of this article is to clarify the setting and the open questions, and also to present some results.

In Section \ref{secspatialrp}, we introduce a model for spatial random permutations in a bounded domain $\Lambda \subset \bbR^d$.
As stated above, the intuition is to suppress permutations with large jumps.
We achieve this by assigning a ``one-body energy'' of the form $\sum_i \xi(x_i - x_{\pi(i)})$ to a given permutation $\pi$ on a finite set $\bsx$.
The one-body potential $\xi$ is nonnegative, spherically symmetric, and typically monotonically increasing, although we will allow more general cases.
In addition, we will introduce ``many-body potentials'' depending on several jumps, as well as a weight for the points $\bsx$.

As usual, the most interesting mathematical structures will emerge in the thermodynamic limit $|\Lambda|, N \to \infty$, where $N$ is the number of points in $\bsx$.
The ambitious approach is to consider and study the infinite volume limit of probability measures.
The limiting measure should be a well-defined joint probability measure on countably infinite (but locally finite) sets $\bsx \subset \bbR^d$, and on permutations of $\bsx$.
To establish such a limit seems fairly difficult;
as an alternative one can settle for constructing an infinite volume measure for permutations only, with a fixed $\bsx$ chosen according to some point process.
We provide a framework for doing so in Section \ref{secinfvol}, and give a natural criterion for 
the existence of the infinite volume limit (it is a generalisation of the one given in \cite{GRU1}).
This criterion is trivially fulfilled if the interaction prohibits jumps greater than a certain finite distance;
however, its verification for the physically most interesting cases remains an open problem. 

Another option for taking the thermodynamic limit is to focus on the existence of the limiting distribution of one special random variable as $|\Lambda|, N \to \infty$.
Motivated by its relevance to Bose-Einstein condensation, our choice of random variable is the probability of the existence of long cycles;
more precisely, we will study the fraction of indices that lie in a cycle of macroscopic length.
The general intuition is that in situations where points are sparse (low density), or where moderately long jumps are strongly discouraged (high temperature), the typical permutation is a small perturbation of the identity map, and there are no infinite cycles.
In Section \ref{secnoinfinitecycles} we give a criterion (that corresponds to low densities, resp.\ high temperatures) for the absence of infinite cycles. 

On the other hand, infinite cycles are usually present for high density.
The density where existence of infinite cycles first occurs is called the {\it critical density}. 
We establish the occurrence of infinite, macroscopic cycles in Section \ref{secinfcycles} for the case where only the one-body potential is present, and where we average over the
point configurations $\bsx$ in a suitable way. 
An especially pleasing aspect of the result is the existence of a simple, {\it exact formula} for the critical density.
It turns out to be nothing else than the critical density of the ideal Bose gas, first computed by Einstein in 1925!
The experienced physicist may shrug this fact off in hindsight.
However, it is a priori not apparent why quantum mechanics should be useful in understanding this problem, and it is fortunate that much progress has been achieved on bosonic systems over the years.
Of direct relevance here is S\"ut\H o's study of the ideal gas \cite{Suto2}, and the work of Buffet and Pul\'e on distributions of occupation numbers \cite{BP}.

Section \ref{secBosegas} is devoted to the relation between models of spatial random permutations and the Feynman-Kac representation of the Bose gas. We are particularly interested in the effect of interactions on the Bose-Einstein condensation. While this question has been largely left to numericians, experts in path-integral Monte-Carlo methods, we expect that weakly interacting bosons can be exactly described by a model of spatial permutations with two-body interactions.
An interesting open problem is to establish this fact rigorously.
Numerical simulations of the model of spatial permutations should be rather easy to perform, and they should help us to understand the phase transition to a Bose condensate.

In Section \ref{secsimplemodel} we simplify the interacting model of Section \ref{secBosegas}. 
It turns out that the largest terms contributing to the interactions between permutation jumps are due to cycles of length $2$.
Retaining this contribution only, we obtain a toy model of interacting random permutations that is simple enough to handle, but which allows to explore some of the effects of interactions on Bose-Einstein condensation.
In particular, we are able to compute the critical temperature exactly.
It turns out to be higher than the non-interacting one and to deviate linearly in the scattering length of the interaction potential.
This is in qualitative agreement with the findings of the physical community \cite{AM,KPS,Kas,NL}.
In addition, we show that infinite cycles occur in our toy model whenever the density is sufficiently high.
However, our condition on the density is not optimal
--- higher than the critical density of our interacting model.
Our condition is not optimal.
We expect the existence of infinite cycles right down to the interacting critical density, but this is yet another open problem.

\medskip
\noindent
{\bf Acknowledgments.}
We are grateful to many colleagues for discussions over a rather long period of time.
In particular, we would like to mention Stefan Adams, Michael Aizenman, Marek Biskup, J\"urg Fr\"ohlich, Daniel Gandolfo, Gian Michele Graf, Martin Hairer,
Roman Koteck\'y, Alain Joye, Elliott Lieb, Bruno Nachtergaele, Charles-\'Edouard Pfister, Suren Poghosyan, Jose Luis Rodrigo, Jean Ruiz, Benjamin Schlein, Robert Seiringer,
Herbert Spohn, Andras S\"ut\H o, Florian Theil, Jochen Voss, Valentin Zagrebnov, and Hans Zessin.
V.B.\ is supported by the EPSRC fellowship EP/D07181X/1.
D.U.\ is supported in part by the grant DMS-0601075 of the US National Science Foundation.

\section{The model in finite volume}
\label{secspatialrp}

Let $\Lambda$ be a bounded open domain in $\bbR^d$, and let $V$ denote its volume (Lebesgue measure).
The state space of our model is the cartesian product
\be
\Omega_{\Lambda,N} = \Lambda^N \times \caS_N,
\ee
where $\caS_N$ is the symmetric group of permutations of \{1,\ldots,N\}.
The state space $\Omega_{\Lambda,N}$ can be equipped with the product $\sigma$-algebra of the Borel $\sigma$-algebra for $\Lambda^N$, and the discrete $\sigma$-algebra for $\caS_N$.
An element $(x_1, \ldots, x_N) \times \pi \in \Omega_{\Lambda,N}$ is viewed as a spatial random permutation 
in the sense that $x_j$ is mapped to $x_{\pi(j)}$ for all $j$. Figure \ref{figpermutation} illustrates this.
\bfig
\centerline{\includegraphics[width=70mm]{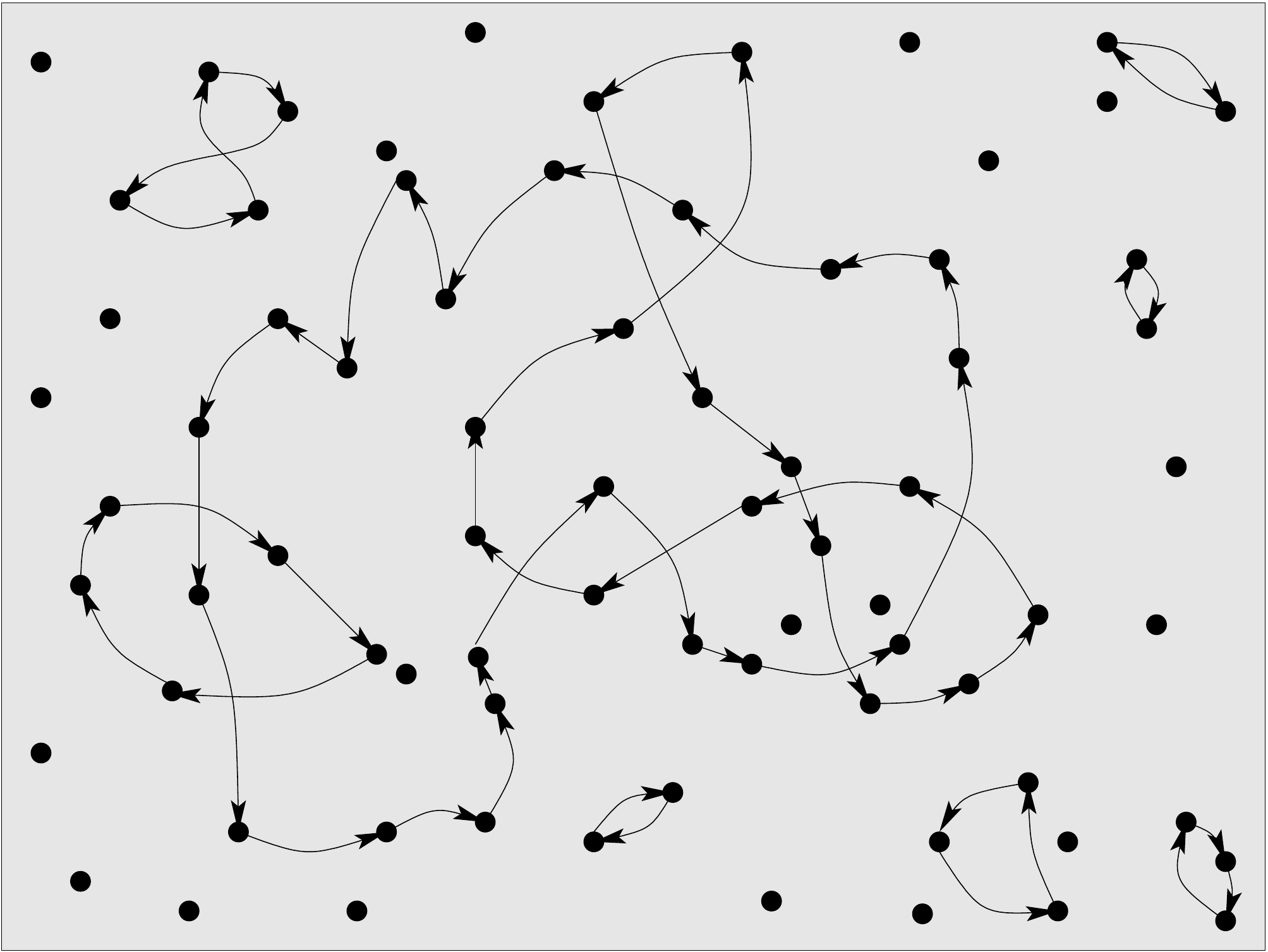}}
\caption{Illustration for a random set of points $\bsx$, and for a permutation $\pi$ on $\bsx$.
Isolated points are sent onto themselves.}
\label{figpermutation}
\efig
The probability measure on $\Omega_{\Lambda,N}$ is obtained in the usual way of statistical mechanics:
a reference measure, in our case the product of Lebesgue measure on $\Lambda^N$ and uniform measure on permutations, is perturbed by a density given by the exponential of a {\it Hamiltonian},
i.e.\ a function $H : \Omega_{\Lambda,N} \to (-\infty,\infty]$.
We will shortly specify the shape of relevant Hamiltonians.

We are interested in properties of permutations rather than positions, and we only consider random variables on $\caS_N$.
We consider two different expectations:
$E_\bsx$, when positions $\bsx \in \Lambda^N$ are fixed; and $E_{\Lambda,N}$, when we average over positions.
For this purpose we introduce the partition functions
\be
\begin{split}
&Y(\bsx) = \sum_{\pi \in \caS_N} \e{-H(\bsx,\pi)}, \\
&Z(\Lambda,N) = \frac1{N!} \int_{\Lambda^N} Y(\bsx) \,\dd \bsx.
\end{split}
\ee
In the last line, $\dd \bsx$ denotes the Lebesgue measure on $\bbR^{dN}$.
The factor $1/N!$ implies that $Z(\Lambda,N) \sim \e{Vq}$ for large $V,N$, and for ``reasonable'' Hamiltonians --- a desirable property in statistical mechanics.
Then, for $\theta : \caS_N \to \bbR$ a random variable on the set of permutations, we define
\be
E_\bsx(\theta) = \frac1{Y(\bsx)} \sum_{\pi\in\caS_N} \theta(\pi) \e{-H(\bsx,\pi)},
\ee
and
\be
\begin{split}
E_{\Lambda,N}(\theta) &= \frac1{Z(\Lambda,N) N!} \int_{\Lambda^N} \dd \bsx \sum_{\pi\in\caS_N} \theta(\pi) \e{-H(\bsx,\pi)} \\
&= \frac1{Z(\Lambda,N) N!} \int_{\Lambda^N} E_\bsx(\theta) Y(\bsx) \dd \bsx.
\label{defexpectation}
\end{split}
\ee

We will be mostly interested in the possible occurrence of long cycles.
Thus we introduce a random variable that measures the {\it density of points in cycles of length between $m$ and $n$}:
\be
\label{defdenscycles}
\bsvarrho_{m,n}(\pi) = \tfrac1V \, \# \bigl\{ i=1,\dots,N : m \leq \ell_i(\pi) \leq n \bigr\}.
\ee
Here, $\ell_i(\pi)$ denotes the length of the cycle that contains $i$;
that is, $\ell_i(\pi)$ is the smallest number $n\geq1$ such that $\pi^{(n)}(i) = i$.
We also have
\be
\bsvarrho_{m,n}(\pi) = \frac1{V} \sum_{i=1}^N \upchi_{[m,n]} \bigl( \ell_i(\pi) \bigr)
\ee
with $\upchi_I$ denoting the characteristic function for the interval $I$.

We denote by $\caR_N$ the space of random variables on $\caS_N$ that are invariant under transpositions.
That is, $\theta \in \caR_N$ satisfies
\be
\theta(\sigma^{-1} \pi \sigma) = \theta(\pi)
\ee
for all $\sigma, \pi \in \caS_N$. Random variables in $\caR_N$ have the useful property that they do not depend on the way the set $\bsx = \{x_1, \ldots, x_N\}$ is labeled.
Instead, they only depend on the set $\bsx$ itself, and in that sense are the most natural quantities to study. 
Notice that $\bsvarrho_{m,n} \in \caR_N$.

We now discuss the form of relevant Hamiltonians.
$H$ is given by the sum
\be
H(\bsx,\pi) = H^{(1)}(\bsx,\pi) + \sum_{k\geq2} H^{(k)}(\bsx,\pi) + G(\bsx),
\ee
where the terms satisfy the following properties.
Let $\bsx = (x_1,\dots,x_N)$.
\begin{itemize}
\item The one-body Hamiltonian $H^{(1)}$ has the form
\be
H^{(1)}(\bsx,\pi) = \sum_{i=1}^N \xi(x_i - x_{\pi(i)}).
\ee
We suppose that $\xi$ is a spherically symmetric function $\bbR^d \to [0,\infty]$, that $\xi(0)=0$, and that $\e{-\xi}$ is integrable.
\item The $k$-body term $H^{(k)} : \Omega_{\Lambda,N} \to \bbR$ can be negative; is has the form
\be
H^{(k)}(\bsx,\pi) = \sumtwo{A \subset \{1,\ldots,N\}}{|A|=k} V \bigl( (x_j, x_{\pi(j)})_{j \in A} \bigr).
\ee
\item The function $G : \bbR^{dN} \to \bbR$ depends on the points only.
It has no effect on the expectation $E_\bsx$, but it modifies the expectation $E_{\Lambda,N}$.
\end{itemize}

We will discuss in Section \ref{secBosegas} the links between spatial random permutations and the quantum Bose gas.
We will see that the physically relevant terms are $\xi(x) = |x|^2/4\beta$, that the interactions are two-body ($H^{(k)}=0$ for $k>2$), and that $G(\bsx) \equiv 0$.
From the mathematical point of view it is interesting to consider a more general setting.
In particular, we can restrict the jumps by setting $\xi(x)=\infty$ for $|x|$ bigger than some cutoff distance $R$.
The effect of $G$ is to modify the typical sets of points.
We can choose it such that $Y(\bsx) \equiv 1$.
We refer to this case as ``Poisson'', since positions are independent of each other, and they are uniformly spread.
The point process for the Bose gas is not Poisson, however.
The fluctuations of the number of points in a subdomain were studied in \cite{LLS}.
They were shown to satisfy a large deviation principle with a rate function that is different than Poisson's.

\section{The model in infinite volume}
\label{secinfvol}

As usual, the most interesting structures emerge in the 
infinite volume limit $V \to \infty$ or, more precisely, the thermodynamic limit $V \to \infty$, 
$N = \rho V$. The easiest way to take this limit is to consider a fixed random variable, e.g.\ 
$\bsvarrho_{a,b}(\pi)$ from (\ref{defdenscycles}), and study its distribution as $V \to \infty$ and $N = \rho V$.
We will indeed do this in Section \ref{secinfcycles};
an advantage of this approach is that we do not have to worry about infinite volume probability measures.
But these infinite volume measures are very interesting objects to study directly, in the same spirit as when constructing infinite volume Gibbs measures.
We advocate this point of view in the present section, and introduce a framework for spatial permutations in unbounded domains.

\subsection{The $\sigma$-algebra}
\label{subs 3.1}

The present and the following subsection contain preparatory results about permutations on $\N$.
We introduce the ``cylinder sets'' $B_{i,j}$ that consist of all permutations where $i \in \N$ is sent to $j \in \N$:
\be
\label{B_xy}
B_{i,j} = \{ \pi \in \caS_\bbN : \pi(i)=j \}.
\ee
Let $\Sigma'$ denote the collection of finite intersections of cylinder sets and their complements.
One can check that it is closed under finite intersections, and also that the difference of two sets is equal to a finite union of disjoint sets.
Such a family of sets is called a {\it semiring} by probabilists.
Semirings are useful because they are easy to build from basic sets, and because premeasures on semirings can be extended to measures by the Carath\'eodory-Fr\'echet theorem.
Let $\Sigma$ be the $\sigma$-algebra generated by the $B_{i,j}$. 

We start by proving a structural lemma that we shall use when extending finite volume measures to an infinite volume one.
For $A_1, A_1', \dots, A_m, A_m' \subset \bbN$, let us define
\be
\label{useful sets}
B^{A_1 \dots A_m}_{A_1' \dots A_m'} = \bigl\{ \pi \in \caS_\bbN : \pi^{-1}(i) \in A_i \text{ and } \pi(i) \in A_i', 1 \leq i \leq m \bigr\}.
\ee
One easily checks that any element of the semiring can be represented by a set of the form \eqref{useful sets}.
Also, the intersection of two such sets satisfies
\be
\label{intersection of sets}
B^{A_1 \dots A_m}_{A_1' \dots A_m'} \Inter B^{C_1 \dots C_m}_{C_1' \dots C_m'} = B^{A_1 \cap C_1 \dots A_m \cap C_m}_{A_1' \cap C_1' \dots A_m' \cap C_m'}.
\ee

\begin{lemma}
\label{nonempty crit}
Let $A_1, A_1', A_2, A_2', \dots$ be finite subsets of $\bbN$.
If
\[
B^{A_1 \dots A_m}_{A_1' \dots A_m'} \neq \emptyset
\]
for all finite $m$, then $\lim_{m\to\infty} B^{A_1 \dots A_m}_{A_1' \dots A_m'} \neq \emptyset$.
\end{lemma}

It is crucial that both $A_i$ and $A_i'$ be finite for all $i$; counter-examples are easily found otherwise.
For instance, choose $A_1 = \bbN$, $A_i = \{i-1\}$ for $i>1$, and $A_i' = \{i+1\}$ for all $i$;
then each finite intersection is non-empty.
The infinite intersection is empty, on the other hand, since there is no possibility left for the preimage of 1.
Similarly, choosing $A_1' = \bbN$, $A_i' = \{i-1\}$ for $i>1$ and $A_i = \{i+1\}$ does not leave a possible image for 1.
These two cases should be kept in mind when reading the proof.

A claim similar to Lemma \ref{nonempty crit} and Theorem \ref{perm meas exist} was proposed in \cite{GRU1}.
The proof there contains a little flaw that is corrected here.

\begin{proof}
We write $B^a_{a'}$ instead of $B^{\{a\}}_{\{a'\}}$, etc...
We have
\be
\label{disj union}
B^{A_1 \dots A_m}_{A_1' \dots A_m'} = \Uniontwo{a_1,\dots,a_m}{a_1',\dots,a_m'} B^{a_1 \dots a_m}_{a_1' \dots a_m'}.
\ee
The union is over $a_1 \in A_1, \dots, a_m' \in A_m'$, with the restriction that $a_i \neq a_j$ and $a_i' \neq a_j'$ for $i \neq j$.
The union is disjoint, $B^{a_1 \dots a_m}_{a_1' \dots a_m'} \neq \emptyset$, and
\be
B^{a_1 \dots a_{m+1}}_{a_1' \dots a_{m+1}'} \subset B^{a_1 \dots a_m}_{a_1' \dots a_m'}.
\ee
Permutations are charaterised by a ``limiting set'', namely
\be
\{\pi\} = B^{a_1 a_2 \dots}_{a_1' a_2' \dots}
\ee
where $\{a_i\}, \{a_i'\}$ are given by
\be
\label{rien d'important}
\pi^{-1}(i) = a_i \qquad \text{and} \qquad \pi(i) = a_i'.
\ee
Conversely, given $\{a_i\}, \{a_i'\}$, $B^{a_1 a_2 \dots}_{a_1' a_2' \dots}$ is either empty, or it contains the permutation $\pi$ that satisfies \eqref{rien d'important}.
We now check that the decomposition \eqref{disj union} yields at least one non-empty limiting set.

The sets $B^{a_1 \dots a_m}_{a_1' \dots a_m'}$ that appear in \eqref{disj union} can be organised as a tree.
The root is $\caS_\bbN$.
The sets with $m=1$ are connected to $\caS_\bbN$, i.e.\ all the sets $B^{a_1}_{a_1'}$, $a_1 \in A_1$ and $a_1' \in A_1'$.
The sets with $m=2$ are connected to those with $m=1$.
Precisely, the sets $B^{a_1 a_2}_{a_1' a_2'}$, $a_2 \in A_2$ and $a_2' \in A_2'$, are connected to $B^{a_1}_{a_1'}$.
And so on...
There are infinitely many vertices, but finitely many vertices at finite distance from the root.

We can select a limiting set as follows.
Let $\ell^{a_1 \dots a_m}_{a_1' \dots a_m'}$ denote the length of the longest path descending from $B^{a_1 \dots a_m}_{a_1' \dots a_m'}$.
For each $m$, there is at least one $\{a_i\}, \{a_i'\}$ such that $\ell^{a_1 \dots a_m}_{a_1' \dots a_m'} = \infty$.
(Otherwise the tree cannot have infinitely many vertices, since the sets $A_1,A_1',\dots$ have finite cardinality.)
Further, there exists $a_{m+1} \in A_{m+1}$ and $a_{m+1}' \in A_{m+1}'$ such that $\ell^{a_1 \dots a_{m+1}}_{a_1' \dots a_{m+1}'} = \infty$.
We can choose an infinite descending path such that $\ell^{a_1 \dots a_m}_{a_1' \dots a_m'}$ is always infinite.
The sets $B^{a_1 \dots a_m}_{a_1' \dots a_m'}$ are not empty for any $m$, so that $B^{a_1 a_2 \dots}_{a_1' a_2' \dots}$ is a non-empty limiting set.
\end{proof}

\subsection{An extension theorem}
\label{subs extension}

We now give a criterion for a set function $\mu$ on $\Sigma'$ to extend to a measure on $\Sigma$.

\begin{theorem} \label{perm meas exist}
Let $\Sigma'$ be the semiring generated by the cylinder sets $B_{i,j}$ in \eqref{B_xy}, and let $\mu$ be an additive set function on $\Sigma'$ with $\mu(\caS_\bbN) < \infty$.
We assume that for all $i \in \bbN$, and all $\eps > 0$, there exists a finite set $A \subset \bbN$ (that depends on $i$ and $\eps$) such that
\be
\label{exist crit}
\sum_{j\in A} \mu(B_{i,j}) > 1-\eps, \qquad \text{and} \qquad \sum_{j\in A} \mu(B_{j,i}) > 1-\eps.
\ee
Then $\mu$ extends uniquely to a measure on $(\caS_{\bbN}, \Sigma)$.
\end{theorem}

If $\mu$ is symmetric with respect to the inversion of permutation, i.e.\ $\mu(B_{i,j}) = \mu(B_{j,i})$, then the two conditions are equivalent to
\be
\sum_{j=1}^\infty \mu(B_{i,j}) = \mu(\caS_\bbN)
\ee
for any fixed $i$.
Since $\caS_\bbN = \cup_j B_{i,j}$, the condition amounts to $\sigma$-additivity for a restricted class of disjoint unions.
Thus \eqref{exist crit} is also a necessary condition.

\begin{proof}
It follows from the assumption of the theorem that for any $\eps>0$, there exist finite sets $C_1,C_2,\dots$ such that, for all $i$,
\be
\mu \bigl( \{ \pi \in \caS_\bbN : \pi^{-1}(i) \in C_i \text{ and } \pi(i) \in C_i \} \bigr) > 1 - 2^{-i-1} \eps.
\ee
Then for any $m$, we have
\be
\mu \bigl( B^{C_1 \dots C_m}_{C_1' \dots C_m'} \bigr) > 1 - \tfrac12 \eps.
\ee
We prove the following property, that is equivalent to $\sigma$-additivity in $\Sigma'$.
For any decreasing sequence $(G_n)$ of sets in $\Sigma'$ such that $\mu(G_n) > \eps$ for all $n$, we have $\lim_n G_n \neq \emptyset$.
As we have observed above, $G_n$ can be written as
\be
G_n = B^{A_{n1} \dots A_{nm}}_{A_{n1}' \dots A_{nm}'}
\ee
for some finite $m$ that depends on $n$.
Without loss of generality, we can suppose that $m \geq n$ (choosing some sets to be $\bbN$ if necessary);
we can actually take $m=n$ (by restricting to a subsequence if necessary).
This allows to alleviate a bit the argument.

We have $G_n \cap G_{n+1} = G_{n+1}$.
Using \eqref{intersection of sets}, we can choose the sets such that $A_{ni} \supset A_{n+1,i}$ and $A_{ni}' \supset A_{n+1,i}'$ for each $i$.
The sets $A_{ni}, A_{ni}'$ may be infinite.
We therefore define the finite sets $D_{ni} = A_{ni} \cap C_i$ and $D_{ni}' = A_{ni}' \cap C_i$.
Then
\be
\begin{split}
\mu \bigl( B^{D_{n1} \dots D_{nn}}_{D_{n1}' \dots D_{nn}'} \bigr) &= \mu \bigl( G_n \inter B^{C_1 \dots C_n}_{C_1 \dots C_n} \bigr) \\
&\geq \mu(G_n) - \mu \bigl( \bigl[ B^{C_1 \dots C_n}_{C_1 \dots C_n} \bigr]^{\rm c} \bigr) \\
&> \tfrac12 \eps.
\end{split}
\ee
The sets $D_{ni}, D_{ni}'$ are finite and decreasing for fixed $i$.
Let
\be
D_i = \lim_n D_{ni}, \qquad D_i' = \lim_n D_{ni}'.
\ee
For any $k$, there exists $n$ large enough such that
\be
B^{D_{n1} \dots D_{nn}}_{D_{n1}' \dots D_{nn}'} \subset B^{D_1 \dots D_k}_{D_1' \dots D_k'} \subset G_k.
\ee
The set in the left side is not empty since it has positive measure.
Thus the set in the middle is not empty for any $k$.
By Lemma \ref{nonempty crit}, its limit as $k \to \infty$ is not empty, which proves that $\lim_k G_k \neq \emptyset$.
\end{proof}

Although Theorem \ref{perm meas exist} does not contain any reference to space, we see in the following section that the spatial structure provides a natural way for (\ref{exist crit}) to be fulfilled in certain cases.

\subsection{Permutation cycles and probability measure}

Here we discuss the connection between the considerations of the two previous subsections, and the spatial structure.
The set of permutations where $i$ belongs to a cycle of length $n$ can be expressed as
\be
B_i^{(n)} = \Union_{j_1, \ldots, j_n} \Inter_{i=1}^n B_{j_{i-1},j_i}
\ee
where $j_1,\dots,j_n$ are distinct integers, and where we set $j_0 \equiv j_n$.
The union is countable and $B_i^{(n)}$ belongs to the $\sigma$-algebra $\Sigma$.
The event where $i$ belongs to an infinite cycle is then
\be
B_i^{(\infty)} = \Bigl[ \Union_{n\geq1} B_i^{(n)} \Bigr]^\compl
\ee
and it also belongs to $\Sigma$.
We can introduce the random variable $\ell_i$ for the length of the cycle that contains $i$.
It can take the value $\infty$.
Since $\ell_i^{-1}(\{n\}) = B_i^{(n)}$ for all $n = 1, 2, \dots, \infty$, we see that $\ell_i$ is measurable.

In general the probability distribution of $\ell_i$ depends on $i$.
Thus we average over points in a large domain.
Let $\bsx \subset \bbR^d$ be a countable set with no accumulation points (that is, if $\Lambda \subset \bbR^d$ is bounded, then $\bsx \cap \Lambda$ is finite).
The elements $x_1,x_2,\dots$ of $\bsx$ can be ordered according to their distance to the origin.
More exactly, we suppose that for any cube $\Lambda$ centered at the origin, there exists $N$ such that $x_i \in \Lambda$ iff $i \leq N$.
Let $V$ be the volume of $\Lambda$.
We introduce the density of points in cycles of length between $m$ and $n$, by
\be
\bsvarrho_{m,n}^{(\Lambda)}(\pi) = \tfrac1{V} \, \# \bigl\{ i = 1,\dots,N : m \leq \ell_i(\pi) \leq n \bigr\}.
\ee
This expression is of course very similar to Eq.\ \eqref{defdenscycles} for the model in finite volume.

Next we define the relevant measure on $\caS_\bbN$.
Let $\caS_N$ be the set of permutations that are trivial for indices larger than $N$:
\be
\caS_N = \{ \pi \in \caS_\bbN : \pi(i)=i \text{ if } i>N \}.
\ee
We define the finite volume probability of a set $B$ in the semiring $\Sigma'$ by
\be
\label{nubsx}
\nu_\bsx^{(\Lambda)}(B) = \frac1{Y(\bsx_\Lambda)} \sum_{\pi \in B \cap \caS_N} \e{-H(\bsx_\Lambda,\pi)}.
\ee
Here, $\bsx_\Lambda = \bsx \cap \Lambda$.
The Hamiltonian $H(\bsx_\Lambda,\pi)$ and the normalisation $Y(\bsx_\Lambda)$ are given by the same expression as in Section \ref{secspatialrp}.

The existence of the thermodynamic limit turns out to be difficult to establish.
If $\bsx$ is a lattice such as $\bbZ^d$, or if $\bsx$ is the realisation of a translation invariant point process, we expect that $\nu_\bsx^{(\Lambda)}(B)$ converges as $V\to\infty$.
We cannot prove such a strong statement, but it follows from Cantor's diagonal argument that there exists a subsequence $(V_n)$ of increasing volumes, such that $\nu_\bsx^{(\Lambda_n)}(B)$ converges for all $B \in \Sigma'$.
(Here, $\Lambda_n$ is the cube of volume $V_n$ centered at 0.)
Thus we have existence of a limiting set function, $\nu_\bsx$, but we cannot garantee its uniqueness.

If $\xi$ involves a cutoff, i.e.\ if $\e{-\xi(x)}$ is zero for $x$ large enough, (\ref{exist crit}) is fulfiled and $\nu_\bsx$ extends to an infinite volume measure thanks to Theorem \ref{perm meas exist}.
But we cannot prove the criterion of the theorem for more general $\xi$, not even the Gaussian.
It is certainly true, though.

For relevant choices of point processes and of permutation measures, we expect that $\lim_{V\to\infty} \bsvarrho_{m,n}^{(\Lambda)}(\pi)$ exists for a.e.\ $\bsx$ and a.e.\ $\pi$.
Let $\bsvarrho_{m,n}$ denote the limiting random variable.
It allows to define the expectation
\be
E(\bsvarrho_{m,n}) = \int\dd\mu(\bsx) \int\dd\nu_\bsx(\pi) \bsvarrho_{m,n}(\pi).
\ee
It would be interesting to obtain properties such as concentration of the distribution of $\bsvarrho_{m,n}$.
The simplest case should be the Poisson point process.

\subsection{Finite vs infinite volume}
\label{subsect finvsinf}

The finite volume setting of Section \ref{secspatialrp} can be rephrased in the infinite volume setting as follows.
Let $\mu^{(\Lambda,N)}$ be the point process on $\bbR^d$ such that
\be
\mu^{(\Lambda,N)}(\dd\bsx) = \begin{cases} \frac{Y(\bsx)}{Z(\Lambda,N) N!} \dd \bsx & \text{if $\bsx \subset \Lambda$ and $|\bsx|=N$,} \\ 0 & \text{otherwise.} \end{cases}
\ee
Next, let $\nu_\bsx^{(\Lambda)}$ be as in (\ref{nubsx}).
For $\Lambda_1 \supset \Lambda_2 \supset \Lambda_3$, let us consider the expectation
\be
\label{defnewexpectation}
E_{\Lambda_1,\Lambda_2,N}(\bsvarrho_{m,n}^{(\Lambda_3)}) = \int\dd\mu^{(\Lambda_1,N)}(\bsx) \int\dd\nu_\bsx^{(\Lambda_2)}(\pi) \bsvarrho_{m,n}^{(\Lambda_3)}.
\ee
This can be compared to Eqs \eqref{defexpectation} and \eqref{defdenscycles}.
Namely, we have
\be
E_{\Lambda,N}(\bsvarrho_{m,n}) = E_{\Lambda,\Lambda,N}(\bsvarrho_{m,n}^{(\Lambda)}).
\ee
It would be interesting to prove that the infinite volume limits in \eqref{defnewexpectation} can be taken separately.
Precisely, we expect that
\be
\lim_{\Lambda\nearrow\bbR^d} E_{\Lambda,\rho|\Lambda|}(\bsvarrho_{m,n}) = \lim_{\Lambda_3 \nearrow \bbR^d} \lim_{\Lambda_2 \nearrow \bbR^d} \lim_{\Lambda_1 \nearrow \bbR^d} E_{\Lambda_1,\Lambda_2,\rho|\Lambda_1|}(\bsvarrho_{m,n}^{(\Lambda_3)}).
\ee

\section{A regime without infinite cycles}
\label{secnoinfinitecycles}

At low density the jumps are very much discouraged and the typical permutations resemble the identity permutation, up to few small cycles here and there.
In this section we give a sufficient condition for the absence of infinite cycles.
We consider the infinite volume framework.
The condition has two parts: (a) a bound on the strength of the interactions; (b) in essence, that the points of $\bsx$ lie far apart compared to the decay length of $\e{-\xi}$.
Here, $B^{(\gamma)}$ denotes the set of permutations where the cycle $\gamma$ is present.

\begin{theorem}
\label{thmfinitecycles}
Let $\bsx \subset \bbR^d$ be a countable set, and let $i$ be an integer.
We assume the following conditions on interactions and on jump factors, respectively.
\begin{itemize}
\item[(a)] There exists $0 \leq s < 1$ such that for all cycles $\gamma = (j_1,\dots,j_n) \subset \bbN$ long enough, with $j_1 = i$, and all permutations $\pi \in B^{(\gamma)}$,
\[
\sumtwo{A \subset \bbN}{A \cap \gamma \neq \emptyset} \Bigl| V \bigl( (x_j,x_{\pi(j)})_{j \in A} \bigr) - V \bigl( (x_j,x_{\overline\pi(j)})_{j \in A} \bigr) \Bigr| \leq s \sum_{j=1}^n \xi(x_j - x_{j-1}),
\]
with $x_0 \equiv x_n$, and where $\overline\pi$ is the permutation obtained from $\pi$ by removing $\gamma$, i.e.
\[
\overline\pi(j) = \begin{cases} \pi(j) & \text{if } j \neq j_1,\dots,j_n, \\ j & \text{if $j=j_k$ for some $k$.} \end{cases}
\]
\item[(b)] With the same $s$ as in {\rm (a)},
\[
\sum_{n\geq1} \sumtwo{\gamma = (j_1,\dots,j_n)}{j_1=i} \prod_{k=1}^n \e{-(1-s) \xi(x_{j_k} - x_{j_{k-1}})} < \infty.
\]
\end{itemize}
Then we have
\[
\lim_{n\to\infty} \lim_{V\to\infty} \sum_{k\geq n} \nu_\bsx^{(\Lambda)}(B_i^{(k)}) = 0.
\]
\end{theorem}

The condition (a) is stated ``for all cycles long enough'', i.e.\ it must hold for all cycles of length $n \geq n_0$, for some fixed $n_0$ that depends on $i$ only.
This weakening of the condition is useful;
many points are involved, they cannot be too close, and $\sum \xi(\cdot)$ in the right side cannot be too small.

If $\nu_\bsx$ extends to a measure on $\Sigma$, then
\be
\lim_{n\to\infty} \lim_{V\to\infty} \sum_{k\geq n} \nu_\bsx^{(\Lambda)}(B_i^{(k)}) = \nu_\bsx(B_i^{(\infty)}),
\ee
and the theorem states that $\nu_\bsx(B_i^{(\infty)}) = 0$.

An open problem is to provide sufficient conditions on the Hamiltonian such that, in the finite volume framework, we have
\[
\lim_{M\to\infty} \lim_{V\to\infty} E_{\Lambda,\rho V}(\bsvarrho_{M,\rho V}) = 0.
\]

\begin{proof}[Proof of Theorem \ref{thmfinitecycles}]
By definition,
\be
\nu_\bsx^{(\Lambda)}(B_i^{(k)}) = \frac1{Y(\bsx_\Lambda)} \sumtwo{\gamma = (j_1,\dots,j_k)}{j_1 = i} \sum_{\pi \in B_N^{(\gamma)}}
\e{-\sum_{l=1}^k \xi(x_{i_l} - x_{i_{l-1}}) - \sum_{A \cap \gamma \neq \emptyset}  V((x_j,x_{\pi(j)})_{j \in A}) - H(\bsx_{\Lambda \setminus \gamma}, \pi)}
\ee
with $B_N^{(\gamma)} = B^{(\gamma)} \cap \caS_N$.
By restricting the sum over permutations, we get a lower bound for the partition function:
\be
Y(\bsx_\Lambda) \geq \sum_{\pi \in B_N^{(\gamma)}} \exp \Bigl\{ - \sum_{A \cap \gamma \neq \emptyset} V((x_j, x_{\overline\pi(j)})_{j \in A}) - H(\bsx_\Lambda, \overline\pi) \Bigr\}.
\ee
Observe that $H(\bsx_{\Lambda \setminus \gamma}, \pi) = H(\bsx_\Lambda, \overline\pi)$.
Combining with the two conditions of the theorem, we obtain
\be
\sum_{k\geq n} \nu_\bsx^{(\Lambda)}(B_i^{(k)}) \leq \sum_{k\geq n} \sumtwo{\gamma = (j_1,\dots,j_k)}{j_1 = i} \prod_{l=1}^n \e{-(1-s) \xi(x_{j_l} - x_{j_{l-1}})}.
\ee
The right side does not depend on $\Lambda$, and it vanishes in the limit $n \to \infty$.
\end{proof}

\section{The one-body model}
\label{secinfcycles}

\subsection{Occurrence of infinite cycles}

The occurrence of infinite cycles can be proved in a large class of models with {\it one-body} Hamiltonians, with the critical density being exactly known!
Here the domain $\Lambda$ is the cubic box of size $L$, volume $V=L^d$.
We fix the density $\rho$ of points, that is, we take $N = \rho V$.

Recall the definition \eqref{defdenscycles} for the random variable $\bsvarrho_{m,n}$ that gives the density of points in cycles of length between $m$ and $n$.
We consider the Hamiltonian
\be
H(\bsx,\pi) = \sum_{i=1}^N \xi_\Lambda(x_i - x_{\pi(i)}),
\ee
where the function $\xi_\Lambda$ is a slight modification of the function $\xi$, defined by the relation
\be
\e{-\xi_\Lambda(x)} = \sum_{y \in \bbZ^d} \e{-\xi(x-Ly)}.
\ee
Our conditions on $\xi$ ensure that the latter sum is finite, and that $\xi_\Lambda$ converges pointwise to $\xi$ as $V\to\infty$.
This technical modification should not be necessary, but it allows to simplify the proof of Theorem \ref{thminfinitecycles} below.
In essence, this amounts to choosing periodic boundary conditions.

Let $C = \int \e{-\xi}$.
We suppose that the Fourier transform of $\e{-\xi(x)}$ is nonnegative, and we denote it $C \e{-\varepsilon(k)}$:
For $k \in \bbR^d$,
\be
\label{defFourier}
C \e{-\varepsilon(k)} = \int_{\bbR^d} \e{-2\pi\ii kx} \e{-\xi(x)} \dd x
\ee
The ``dispersion relation'' $\varepsilon(k)$ always satisfies
\begin{itemize}
\item $\varepsilon(0) = 0$; $\varepsilon(k) > a|k|^2$ for small $k$
(indeed, the Laplacian of $\e{-\varepsilon(k)}$ would otherwise be zero at $k=0$; then $\int |x|^2 \e{-\xi(x)} \dd x = 0$, which is absurd).
\item $\varepsilon(k) > 0$ uniformly in $k$ away from zero;
\item $\int \e{-\varepsilon(k)} \dd k = C^{-1} < \infty$.
\end{itemize}

Among many examples of functions that satisfy the conditions above, let us mention several important cases.
\begin{enumerate}
\item The Gaussian:
\be
\e{-\xi(x)} = \e{-|x|^2 / 4\beta}, \quad\quad \varepsilon(k) = 4\pi^2 \beta |k|^2, \quad\quad C = (4\pi\beta)^{d/2}.
\ee
\item In dimension $d=3$, the exponential:
\be
\e{-\xi(x)} = \e{-|x| / \beta}, \quad\quad \varepsilon(k) = 2 \log [1 + (2\pi\beta|k|)^2], \quad\quad C = 8\pi / \beta^3.
\ee
\item In $d=3$, a sufficient condition for positive Fourier transform is that
\be
\tfrac1r \bigl( \e{-\xi(r)} \bigr)''
\ee
be monotone decreasing;
here, $\xi$ depends on $r=|x|$ only.
See e.g.\ \cite{HS} and references therein.
Thus there exist functions $\e{-\xi}$ with compact support and positive Fourier transform.
\item In $d=1$,
\be
\e{-\xi(x)} = (|x|+1)^{-3/2}.
\ee
Its Fourier transform is positive by Lemma \ref{lemposFourier}.
It can be checked that $\varepsilon(k) \sim |k|^{1/2}$ for small $k$, so that its critical density is finite.
\end{enumerate}

We define the {\it critical density} by
\be
\label{critdens}
\rho_{\rm c} = \int_{\bbR^d} \frac{\dd k}{\e{\varepsilon(k)} - 1}.
\ee
The critical density is finite for $d\geq3$, but it can be infinite in $d=1,2$.
The following theorem claims that infinite cycles are present for $\rho > \rho_{\rm c}$ only, and that they are macroscopic.
It extends results of S\"ut\H o for the ideal Bose gas \cite{Suto2}, which corresponds to the Gaussian case $\xi(x) \sim |x|^2$.
The theorem also applies when $\rho_{\rm c} = \infty$.

\begin{theorem}
\label{thminfinitecycles}
Let $\xi$ satisfy the assumptions above. Then for all $0<a<b<1$, and all $s\geq0$,
\[
\begin{split}
&{\rm (a)} \quad \lim_{V\to\infty} E_{\Lambda, \rho V}(\bsvarrho_{1,V^a}) = \begin{cases} \rho & \text{if } \rho \leq \rho_{\rm c}; \\ \rho_{\rm c} & \text{if } \rho \geq \rho_{\rm c}; \end{cases} \\
&{\rm (b)} \quad \lim_{V\to\infty} E_{\Lambda, \rho V}(\bsvarrho_{V^a,V^b}) = 0; \\
&{\rm (c)} \quad \lim_{V\to\infty} E_{\Lambda, \rho V}(\bsvarrho_{V^b,sV}) = \begin{cases} 0 & \text{if } \rho \leq \rho_{\rm c}; \\ s & \text{if } 0 \leq s \leq \rho-\rho_{\rm c}, \\
\rho - \rho_{\rm c} & \text{if } 0 \leq \rho-\rho_{\rm c} \leq s. \end{cases}
\end{split}
\]
\end{theorem}

The rest of this section is devoted to the proof of this theorem.

\subsection{Fourier representation for spatial permutations}

The first step is to reformulate the problem in the Fourier space.
In the Gaussian case, $\xi(x) = |x|^2$, this is traditionally done using the Feynman-Kac formula and 
unitary transformations of Hilbert spaces.
There is no Feynman-Kac formula for our more general setting, but we can directly use the Fourier transform.
This actually simplifies the situation.

Here and in the sequel, we use the definition \eqref{defFourier} for the Fourier transform.
Let $\Lambda$ be the unit cube, i.e.\ we fix the length to $L=1$.
For $f \in L^1(\bbR^d)$, let $f_\Lambda(x) = \sum_{y \in \bbZ^d} f(x+y)$; notice that $f_\Lambda \in L^1(\Lambda)$.

\begin{lemma}
\label{lemFourierconvol}
Let $f \in L^1(\bbR^d)$.
For all $n=1,2,\dots$, we have
\[
\int_{\Lambda^n} \dd x_1 \dots \dd x_n \prod_{i=1}^n f_\Lambda(x_i-x_{i-1}) = \sum_{k \in \bbZ^d} \widehat f(k)^n.
\]
(By definition, $x_0=x_n$.)
\end{lemma}

\begin{proof}
The $n$-th convolution of the function $f_\Lambda$ with itself satisfies
\be
(f_\Lambda^{*n})(x) = \int \dd x_1 \dots \dd x_{n-1} f_\Lambda(x_1) f_\Lambda(x_2-x_1) \dots f_\Lambda(x-x_{n-1}).
\ee
The product $\prod f_\Lambda(x_i-x_{i-1})$ is translation invariant, so that
\be
\begin{split}
\int_{\Lambda^n} \dd x_1 \dots \dd x_n \prod_{i=1}^n f_\Lambda(x_i-x_{i-1}) &= \int_{\Lambda^n} \dd x_1 \dots \dd x_n f_\Lambda(x_1) f_\Lambda(x_2-x_1) f_\Lambda(-x_{n-1}) \\
&= f_\Lambda^{*n}(0) \\
&= \sum_{k \in \bbZ^d} \widehat{f_\Lambda^{*n}}(k) \\
&= \sum_{k \in \bbZ^d} \widehat f(k)^n.
\end{split}
\ee
The hat symbol in the third line denotes the $L^2(\Lambda)$-Fourier transform; but in the fourth line, it denotes the $L^2(\bbR^d)$-Fourier transform.
\end{proof}

We will use a corollary of this lemma.

\begin{corollary}
\label{corFourier}
Let $\Lambda$ be a cube of size $L$, $\Lambda^* = (\frac1L \bbZ)^d$ be the dual space, and let $f_\Lambda(x) = \sum_{y \in \bbZ^d} f(x+Ly)$.
Then for any permutation $\pi \in \caS_N$ we have:
\[
\int_{\Lambda^N} \dd x_1 \dots \dd x_N \prod_{i=1}^N f_\Lambda(x_i - x_{\pi(i)}) = \sum_{k_1, \dots, k_N \in \Lambda^*} \prod_{i=1}^N \delta_{k_i, k_{\pi(i)}} \prod_{i=1}^N \widehat f(k_i).
\]
\end{corollary}

\begin{proof}
It is enough to consider the case $L=1$, the general case can be obtained by scaling.
The multiple integral factorises according to the cycles of the permutation.
Using Lemma \ref{lemFourierconvol} for each cycle, we get the result.
\end{proof}

We are now in position to reformulate the problem in the Fourier space.
By Corollary \ref{corFourier}, we have the relation
\be \label{k expect}
E_{\Lambda,N}(\theta) = \frac1{Z'(\Lambda,N) N!} \sum_{\pi\in\caS_N} \theta(\pi) \sum_{k_1,\dots,k_N \in \Lambda^*} \e{-\sum_{i=1}^N \varepsilon(k_i)} \prod_{i=1}^N \delta_{k_i,k_{\pi(i)}}
\ee
with $Z'(\Lambda,N) = C^{-N} Z(\Lambda,N)$.

A simpler expression than (\ref{k expect}) is available for random variables invariant under transposition, i.e.\ $\theta \in \caR_N$. To obtain it, we introduce the set $\caN_\Lambda$ of ``occupation numbers'' on $\Lambda^*$.
An element $\bsn \in \caN_\Lambda$ is a sequence of integers $\bsn = (n_k)$, $n_k = 0,1,2,\dots$, indexed by $k \in \Lambda^*$.
We denote by $\caN_{\Lambda,N}$ the set of occupation numbers with total number $N$:
\be
\caN_{\Lambda,N} = \Bigl\{ \bsn \in \caN_\Lambda : \sum_{k \in \Lambda^*} n_k = N \Bigr\}.
\ee
Let $\bsk = (k_1,\dots,k_N)$ be an $N$-tuple of elements of $\Lambda^*$.
We can assign an element $\bsn = \bsn(\bsk)$ in $\caN_{\Lambda,N}$ by defining, for each $k \in \Lambda^*$,
\be
n_k = \#\{ i=1,\dots,N : k_i = k \}.
\ee
In other words, $n_k$ is the number of occurrences of the vector $k$ in $\bsk$.
The map $\bsk \mapsto \bsn(\bsk)$ is onto but not one-to-one; the number of $\bsk$'s that are sent to a given $\bsn$ is equal to
\[
N! \Big/ \prod_{k\in\Lambda^*} n_k!.
\]

We now introduce probabilities for Fourier modes, permutations, and occupation numbers.
The sample space is $(\Lambda^*)^N \times \caS_N$; it is discrete, and we consider the discrete $\sigma$-algebra.
When taking probabilities, we write $\bsk \times \pi$ for $\{(\bsk,\pi)\}$; $\bsk$ for $\{\bsk\} \times \caS_N$; $\bsn$ for $\{\bsk : \bsn(\bsk)=\bsn\} \times \caS_N$ and $\pi$ for $(\Lambda^*)^N \times \{\pi\}$.
We define the probability $P_{\Lambda,N}$ by
\be
\label{probF1}
P_{\Lambda,N} \bigl( \bsk \times \pi \bigr) = \begin{cases} \frac1{Z'(\Lambda,N) N!} \e{-\sum_{i=1}^N \varepsilon(k_i)} & \text{if $k_i = k_{\pi(i)}$ for all $i$}, \\ 0 & \text{otherwise.} \end{cases}
\ee
Here, $\bsk = (k_1,\dots,k_N)$.
Summing over permutations, we get
\be
P_{\Lambda,N}(\bsk) = \frac1{Z'(\Lambda,N) N!} \e{-\sum_{k\in\Lambda^*} \varepsilon(k) n_k} \prod_{k\in\Lambda^*} n_k!
\ee
with $(n_k) = \bsn(\bsk)$.
Finally, summing over Fourier modes that are compatible with occupation numbers yields
\be
\label{probF3}
P_{\Lambda,N}(\bsn) = \frac1{Z'(\Lambda,N)} \e{-\sum_{k\in\Lambda^*} \varepsilon(k) n_k}.
\ee
It follows that the partition function $Z'(\Lambda,N)$ can be expressed using occupation numbers as
\be
Z'(\Lambda,N) = \sum_{\bsn \in \caN_{\Lambda,N}} \e{-\sum_{k\in\Lambda^*} \varepsilon(k) n_k}.
\ee

These definitions allow to express $E_{\Lambda,N}(\theta)$ in an illuminating form.

\begin{lemma}
\label{lemkeyFourierproperty}
For all $\theta \in \caR_N$,
\[
E_{\Lambda,N}(\theta) = \sum_{\bsn \in \caN_{\Lambda,N}} P_{\Lambda,N}(\bsn) \sum_{\pi \in \caS_N} \theta(\pi) P_{\Lambda,N}(\pi | \bsk),
\]
with $\bsk$ any $N$-tuple of Fourier modes that is compatible with $\bsn$, i.e.\ such that $\bsn(\bsk)=\bsn$.
\end{lemma}

\begin{proof}
Using the definitions \eqref{probF1}--\eqref{probF3}, the expectation (\ref{k expect}) of $\theta$ can be written as
\be
\begin{split}
E_{\Lambda,N}(\theta) &= \sum_{\bsk \in (\Lambda^*)^N} \sum_{\pi \in \caS_N} \theta(\pi) P_{\Lambda,N}(\bsk \times \pi) \\
&= \sum_{\bsk \in (\Lambda^*)^N} P_{\Lambda,N}(\bsk) \sum_{\pi \in \caS_N} \theta(\pi) P_{\Lambda,N}(\pi | \bsk).
\end{split}
\ee
The latter sum does not depend on the ordering of the $k_i$'s --- it depends only on occupation numbers
(notice that $P_{\Lambda,N}(\pi | \sigma(\bsk)) = P_{\Lambda,N}(\sigma^{-1} \pi \sigma | \bsk)$ for all $\sigma \in \caS_N$).
The lemma follows from \eqref{probF3}.
\end{proof}

\begin{lemma}
\label{lemexpcycles}
\[
\sum_{\pi\in\caS_N} \bsvarrho_{m,n}(\pi) P_{\Lambda,N}(\pi | \bsk) = \frac1{V} \sum_{k\in\Lambda^*} \begin{cases} n-m+1 & \text{if } 1 \leq m \leq n \leq n_k \\ n_k - m + 1 & \text{if } 1 \leq m \leq n_k \leq n \\ 0 & \text{if } n_k < a \leq b. \end{cases}
\]
\end{lemma}

\begin{proof}
It follows from \eqref{probF1} that, given $\bsk$, permutations are uniformly distributed (over compatible permutations).
That is,
\be
P_{\Lambda,N}(\pi | \bsk) = \begin{cases} 1 / \prod_{k\in\Lambda^*} n_k! & \text{if $k_{\pi(i)} = k_i$ for all $i$}, \\ 0 & \text{otherwise.} \end{cases}
\ee
Given $\bsk=(k_1,\dots,k_N)$, a permutation $\pi$ that leaves it invariant can be decomposed into a collection $(\pi_k)$, $\pi_k \in \caS_{n_k}$, of permutations for each Fourier mode, namely
\be
\sum_{\pi \in \caS_N} \bsvarrho_{m,n}(\pi) P_{\Lambda,N}(\pi | \bsk) = \biggl( \prod_{k \in \Lambda^*} \sum_{\pi_k \in \caS_{n_k}} \frac1{n_k!} \biggr) \bsvarrho_{m,n} \bigl( (\pi_k)_{k \in \Lambda^*} \bigr).
\ee
In addition, we have
\be
\bsvarrho_{m,n} \bigl( (\pi_k)_{k\in\Lambda^*} \bigr) = \frac1{V} \sum_{k \in \Lambda^*} N_{m,n}(\pi_k),
\ee
with $N_{m,n}(\pi)$ being the number of indices that belong to cycles of length between $m$ and $n$.
We obtain
\be
\sum_{\pi \in \caS_N} \bsvarrho_{m,n}(\pi) P_{\Lambda,N}(\pi | \bsk) = \frac1{V} \sum_{k\in\Lambda^*} \frac1{n_k!} \sum_{\pi_k \in \caS_{n_k}} N_{m,n}(\pi_k).
\ee
We see that modes have been decoupled;
further, we only need to average $N_{m,n}(\pi)$ over {\it uniform} permutations.
One easily checks that the probability for an index to belong to a cycle of length $\ell$, with $N$ indices, is equal to $1/N$ for any $\ell$.
Then
\be
\frac1{N!} \sum_{\pi \in \caS_N} N_{m,n}(\pi) = n-m+1,
\ee
for integers $1 \leq m \leq n \leq N$.
The lemma follows.
\end{proof}

We have all elements for the proof of Theorem \ref{thminfinitecycles}.

\begin{proof}[Proof of Theorem \ref{thminfinitecycles}]
We introduce a set $A_\eta$ of occupation numbers that is typical.
With $\rho_0 = \max(0, \rho-\rho_{\rm c})$, let
\be
\label{defAeta}
A_\eta = \biggl\{ \bsn \in \caN_{\Lambda,N} : \Bigl| \frac{n_0}{V} - \rho_0 \Bigr| < \eta ;
\sum_{0<|k|<V^{-\eta}} n_k < \eta V ; n_k < V^{3\eta} \text{ for all } |k| \geq V^{-\eta} \biggr\}.
\ee
Notice that, for any $\bsn \in A_\eta$,
\be
\label{nombredemodesgrands}
\min(\rho,\rho_{\rm c}) V - 2\eta V \leq \sum_{|k| \geq V^{-\eta}} n_k \leq \min(\rho,\rho_{\rm c}) V + \eta V.
\ee
It is proved in Proposition \ref{proptypoccnum} that $P_{\Lambda,N}(A_\eta) \to 1$ as $V,N \to \infty$, for all $\eta>0$.
Together with Lemmas \ref{lemkeyFourierproperty} and \ref{lemexpcycles}, we obtain, with $N = \rho V$,
\be
\label{voilaunejolieequation}
\lim_{V\to\infty} E_{\Lambda,\rho V}(\bsvarrho_{m,n}) = \lim_{V\to\infty} \sum_{\bsn \in A_\eta} P_{\Lambda,N}(\bsn) \Bigl[ \bsvarrho_{m,n}^{(0)}(\bsn) + \bsvarrho_{m,n}^{(1)}(\bsn) + \bsvarrho_{m,n}^{(2)}(\bsn) \Bigr].
\ee
We partitioned $\Lambda^*$ into the disjoint union of $M^{(0)} = \{0\}$, $M^{(1)} = \{k : 0<|k|<V^{-\eta}\}$, and $M^{(2)} = \{k : |k| \geq V^{-\eta}\}$, and we introduced
\be
\bsvarrho_{m,n}^{(i)}(\bsn) = \frac1{V} \sum_{k \in M^{(i)}} \Bigl[ (n-m+1) \upchi_{[n,\infty)}(n_k) + (n_k - m + 1) \upchi_{[m,n)}(n_k) \Bigr].
\ee

First, we note that $\bsvarrho_{m,n}^{(1)}(\bsn) < \eta$ for all $m,n$, so this term does not contribute in \eqref{voilaunejolieequation}.
Next, assuming that $3\eta < a < b$, we have
\be
\min(\rho,\rho_{\rm c}) - 2\eta \leq \bsvarrho_{1,V^a}^{(2)}(\bsn) = \frac1{V} \sum_{|k| \geq V^{-\eta}} n_k \leq \min(\rho,\rho_{\rm c}) + \eta.
\ee
We used \eqref{nombredemodesgrands}.
In addition, we observe that $\bsvarrho_{V^a,V^b}^{(2)}(\bsn)$ and $\bsvarrho_{V^b,sV}^{(2)}(\bsn)$ are zero for $\bsn \in A_\eta$.
Let us turn to $\bsvarrho_{m,n}^{(0)}$.
We have
\be
\bsvarrho_{1,V^a}^{(0)}(\bsn) + \bsvarrho_{V^a,V^b}^{(0)}(\bsn) < V^{b-1},
\ee
so these terms do not contribute in \eqref{voilaunejolieequation}.
Finally, we have
\be
\bsvarrho_{V^b,sV}^{(0)}(\bsn) = \begin{cases} 0 & \text{if } \rho \leq \rho_{\rm c}, \\
\frac{n_0}{V} - \frac{V^b+1}{V} & \text{if } \rho \geq \rho_{\rm c} \text{ and } n_0 \leq sV, \\
s - \frac{V^b+1}{V} & \text{if } \rho \geq \rho_{\rm c} \text{ and } n_0 \geq sV. \end{cases}
\ee
Inserting these informations in \eqref{voilaunejolieequation} yields the theorem.
\end{proof}

\section{The quantum Bose gas}
\label{secBosegas}

Einstein understood in 1925 that non-interacting bosons undergo a phase transition where a single particle state becomes macroscopically occupied.
Ever since the Bose gas has stirred the interest of physicists and mathematical physicists.
The first question that theoreticians needed to resolve was whether Bose-Einstein condensation also occurs in interacting systems, and what is the order parameter.
Feynman introduced in 1953 what is now refered to as the Feynman-Kac representation \cite{Fey}.
It involves ``space-time'' trajectories and permutations, and Feynman emphasised the r\^ole played by long cycles.
The correct order parameter, called ``off-diagonal long-range order'', was proposed shortly afterwards by Penrose and Onsager \cite{PO}.
But cycles did not disappear, and it was suggested in 1987 that {\it winding cycles} are related to superfluidity \cite{PC}.
S\"ut\H o made precise the notion of infinite cycles and he showed that it is equivalent to Bose-Einstein condensation in the ideal gas \cite{Suto1,Suto2}.
An exact formula made this relation more explicit \cite{Uel1}.
Recently, Boland and Pul\'e obtained puzzling results for the infinite-hopping Bose-Hubbard model \cite{BP2}:
Infinite cycles are present iff there is Bose-Einstein condensation, but the density of particles in infinite cycles is {\it strictly greater} than that in the condensate.

There are many open questions, of interest to physicists and mathematicians.
The relationship between Bose-Einstein condensation, superfluidity, and infinite cycles, has yet to be clarified.
The ideal gas can condense (if $d\geq3$) but it is never superfluid.
Interacting bosons in one or two dimensions do not condense but they may be superfluid.
Infinite cycles seem related to Bose-Einstein condensation.
It was argued in \cite{Uel2}, however, that infinite cycles may be present in a solid at low temperature, so they do not automatically imply the existence of a condensate (nor a superfluid).
On the other hand, it is expected that interacting bosons in three dimensions have a transition to a Bose condensate and a superfluid, and that this transition takes place at the same critical temperature.
We also expect the critical temperature for infinite cycles to be the same.
It is therefore of physical relevance to consider infinite cycles in the Feynman-Kac representation of the Bose gas.
We discuss in Subsection \ref{sec FKrep} the relation between the Bose gas and models of spatial permutations.
The latter should help us understand the effects of interactions on the critical temperature.
This is discussed in Subsection \ref{sec discussion}.

The Bose gas has also been the object of many interesting mathematical studies.
We only refer to \cite{ZB} for a discussion of Bogoliubov theory and related work, and to \cite{LSSY} for remarkable results on the ground state of the interacting gas.

\subsection{Feynman-Kac representation of the Bose gas}
\label{sec FKrep}

The state space for a system of $N$ identical bosons in a cube $\Lambda \subset \bbR^d$ (size $L$, volume $V=L^d$) is the Hilbert space $L_{\rm sym}^2(\Lambda^N)$ of complex square-integrable symmetric functions on $\Lambda^N$.
The Hamiltonian is the Schr\"odinger operator
\be
H = -\sum_{i=1}^N \Delta_i + \sum_{1 \leq i,j < N} U(x_i - x_j).
\ee
Here, $\Delta_i$ denotes the $d$-dimensional Laplacian for the $i$-th variable, and $U(x_i - x_j)$ is a multiplication operator that represents the interaction between particles $i$ and $j$.
We choose the self-adjoint extension for periodic boundary conditions.

In order to get cycles, and following \cite{Suto1}, we work in the Hilbert space $L^2(\Lambda^N)$ and we consider the unitary representation of $\caS_N$.
That is, let $U_\pi$ denote the unitary operator
\be
U_\pi f(x_1,\dots,x_N) = f(x_{\pi(1)},\dots,x_{\pi(N)}).
\ee
If $\theta$ is a function of permutations, we can define its expectation by
\be
\langle \theta \rangle_{\Lambda,N} = \frac1{Z(\Lambda,N) N!} \sum_{\pi \in \caS_N} \theta(\pi) \Tr U_\pi \e{-\beta H}.
\ee

The Feynman-Kac formula expresses the operator $\e{-\beta H}$ as an integral operator, whose kernel is given by
\bm
K(x_1,\dots,x_N;y_1,\dots,y_N) = \int \dd W_{x_1 y_1}^{2\beta}(\omega_1) \dots \int \dd W_{x_N y_N}^{2\beta}(\omega_N) \\
\exp\Bigl\{ -\sum_{1 \leq i<j \leq N} \int_0^{2\beta} U( \omega_i(s) - \omega_j(s) ) \dd s \Bigr\}.
\end{multline}
Here, the Wiener measure $W_{xy}^\beta$ describes a Brownian bridge between $x$ and $y$ in time $\beta$, with periodic boundary conditions.
We refer to \cite{Gin} for an excellent introduction to the subject.
We have
\be
\label{Wienernormalisation}
\int \dd W_{xy}^\beta(\omega) = \sum_{z\in\bbZ^d} g_\beta(x-y+Lz) \equiv g_\beta^{(\Lambda)}(x-y),
\ee
where $g_\beta$ denotes the normalised Gaussian function
\be
g_\beta(x) = \frac1{(2\pi\beta)^{d/2}} \e{-|x|^2 / 2\beta}.
\ee
The sum over $z$ in \eqref{Wienernormalisation} accounts for periodic boundary conditions.
For large $L$ the functions $g_\beta^{(\Lambda)}$ and $g_\beta$ are almost identical.
If $f$ is a function $\bbR^{dn} \to \bbR$, and $0<t_1<\dots<t_n<\beta$ are ordered ``times'', we have
\bm
\int\dd W^\beta_{xy}(\omega) f \bigl( \omega(t_1), \dots, \omega(t_n) \bigr) = \sum_{z\in\bbZ^d} \int_{\bbR^{dn}} \dd x_1 \dots \dd x_n \\
g_{t_1}^{(\Lambda)}(x-x_1) g_{t_2-t_1}^{(\Lambda)}(x_2-x_1) \dots g_{\beta-t_n}^{(\Lambda)}(y-x_n) f(x_1,\dots,x_n).
\end{multline}
By the Feynman-Kac formula, the partition function is
\be
Z(\Lambda,N) = \Tr_{L^2_{\rm sym}(\Lambda^N)} \e{-\beta H} = \frac1{N!} \sum_{\pi\in\caS_N} \int_{\Lambda^N} \dd x_1 \dots \dd x_N K(x_1, \dots, x_N; x_{\pi(1)}, \dots, x_{\pi(N)}).
\ee
The sum over permutations is present because we work in the symmetric subspace.
The expectation of observables on permutations can be expressed as
\be
\langle \theta \rangle_{\Lambda,N} = \frac1{Z(\Lambda,N) N!} \sum_{\pi\in\caS_N} \theta(\pi) \int_{\Lambda^N} \dd x_1 \dots \dd x_N \, K(x_1, \dots, x_N; x_{\pi(1)}, \dots, x_{\pi(N)}).
\ee
The connection to the model of spatial permutations is immediate in absence of interactions ($U(x) \equiv 0$):
$\langle \theta \rangle_{\Lambda,N} = E_{\Lambda,N}(\theta)$ with $\xi(x) = |x|^2 / 4\beta$, $H^{(k)} = 0$ for $k\geq2$, and $G(\bsx) \equiv 0$.

\subsection{Discussion: Relevant interactions for spatial permutations}
\label{sec discussion}

The models of spatial random permutations retain some of the features of the quantum Bose gas in the Feynman-Kac representation, but not all of them.
The interactions between quantum particles translate into many-body interactions for permutations.
However, an expansion reveals that to lowest order the interaction is two-body;
precisely, the interaction between jumps $x \mapsto y$ and $x' \mapsto y'$ is given by
\be
\label{LaFormule}
V(x,y,x',y') = \int \bigl[ 1 - \e{-\frac14 \int_0^{4\beta} U(\omega(s)) \dd s} \bigr] \dd\widehat W_{x-x',y-y'}^{4\beta}(\omega)
\ee
for $x \neq y$.
Here, $\widehat W^t_{x,y} = g_t^{-1}(x-y) W_{x,y}^t$ is a normalised Wiener measure, $\int\dd\widehat W^t_{x,y} = 1$.
If $U$ consists of a hard-core potential of radius $a$, we notice that $V(x,y,x',y')$ is equal to the probability that a Brownian bridge, from $x-x'$ to $y-y'$, intersects the ball of radius $a$ centered at 0.

The computation of \eqref{LaFormule} can be found in \cite{Uel3};
it is partly justified by a cluster expansion, although a rigorous result is still lacking.
We expect that the critical temperature for the occurrence of infinite cycles is {\it the same} for $\langle \cdot \rangle_{\Lambda,N}$ and $E_{\Lambda,N}(\cdot)$ to lowest order in the strength of the interaction potential.

An important question about Bose systems concerns the effect of interactions on the critical temperature.
Over the years physicists gave several, conflicting answers.
But a consensus has recently emerged in the physics literature, mostly from path-integral Monte-Carlo numerical studies.
The critical temperature $T_{\rm c}^{(a)}$, as a function of the ``scattering length'' $a$ of the interaction potential (see e.g.\ \cite{LSSY} for the definition), behaves in three dimensions as
\be
\label{crittempa}
\frac{T_{\rm c}^{(a)} - T_{\rm c}^{(0)}}{T_{\rm c}^{(0)}} = c \rho^{1/3} a + o(\rho^{1/3} a),
\ee
with $c \approx 1.3$. See \cite{AM,KPS,Kas,NL} and references therein.
The model of spatial random permutations should give the exact correction to the critical temperature, i.e.\ the correct constant $c$ in \eqref{crittempa}.

It is worth discussing the effects of interactions on spatial permutations in detail.
They both modify the point process and the measure on permutations.
Let $\mu^{(a)}$ denote the point process for an interaction potential of scattering length $a$, and $\nu_\bsx^{(a)}$ the measure on permutations.
We can assign different parameters to these two measures, so as to have the expectation
\be
E^{(a,a')}(\theta) = \int\dd\mu^{(a)}(\bsx) \int\dd\nu_\bsx^{(a')}(\pi) \theta(\pi).
\ee
Let us fix the particle density, and let $T_{\rm c}^{(a,a')}$ denote the critical temperature for the occurrence of infinite cycles.
Of course, $T_{\rm c}^{(a)} = T_{\rm c}^{(a,a)}$.
For small $a$ we should have
\be
\frac{T_{\rm c}^{(a)} - T_{\rm c}^{(0)}}{T_{\rm c}^{(0)}} = \frac{T_{\rm c}^{(a,0)} - T_{\rm c}^{(0)}}{T_{\rm c}^{(0)}} + \frac{T_{\rm c}^{(0,a)} - T_{\rm c}^{(0)}}{T_{\rm c}^{(0)}} + o(a).
\ee
The first question is whether both corrections are linear in $a$;
if not, we could happily dismiss one.
This formula may be useful in numerical simulations.
The second term in the right side involves the point process for the ideal gas, while permutation jumps are interacting.
It can be easily calculated numerically since the set of permutations is discrete.
To know the corresponding linear behaviour would already provide some information.
The first term involves the positions of particles, and it may be difficult to generate a typical realisation of the points.
In any case, the random permutation approach with Eq.\ \eqref{LaFormule} is much easier than path-integral Monte-Carlo simulations, and it should yield the same result to lowest order.

\section{A simple model of spatial random permutations with interactions}
\label{secsimplemodel}

In the preceding section we have discussed a two-body interaction that is exactly related to the quantum Bose gas.
Rigorous results seem difficult to get, though.
In this section we simplify the interaction so as to retain only the largest contribution.
Our approximation is not exact, but the model may be of interest as an effective model, and it is exactly solvable.

\subsection{Approximation and definition of the model}

It helps to think of particles as describing Brownian motions, and to interact whenever they cross each other's paths.
Clearly, particles that belong to the same 2-cycle interact a lot.
Our approximation consists in retaining only these interactions, and in neglecting the rest.
Thus we consider the Hamiltonian
\be
\label{Ham approx}
\tilde H(\bsx,\pi) = \frac1{4\beta} \sum_{i=1}^N |x_i - x_{\pi(i)}|^2 + \sumtwo{1\leq i<j\leq N}{\pi(i)=j, \pi(j)=i} V(x_i,x_j,x_j,x_i).
\ee
The potential $V(\cdot)$ is the exact jump interaction given in \eqref{LaFormule}.
To lowest order in the scattering length, we find that \cite{BU2}
\be
\label{V premier ordre}
V(x,y,y,x) = \frac{2a}{|x-y|} + O(a^2).
\ee
If $\theta$ is a random variable that depends only on permutations, its expectation is given by
\be
\label{expectation rv}
E_{\Lambda,N}(\theta) = \frac1{Z(\Lambda,N) N!} \sum_{\pi\in\caS_N} \theta(\pi) \int_{\Lambda^N} \dd\bsx \e{-\tilde H(\bsx,\pi)}.
\ee
This allows to simplify the model \eqref{Ham approx} further, without additional approximation.
Namely, we introduce the simpler Hamiltonian
\be
H^{(\alpha)}(\bsx,\pi) = \frac1{4\beta} \sum_{i=1}^N |x_i - x_{\pi(i)}|^2 + \alpha N_2(\pi),
\ee
with $N_2(\pi)$ denoting the number of 2-cycles in the permutation $\pi$.
Expectations with $\tilde H$ and $H^{(\alpha)}$ are identical provided
\be
\int_{\Lambda^N} \dd\bsx \e{-\tilde H(\bsx,\pi)} = \int_{\Lambda^N} \dd\bsx \e{-H^{(\alpha)}(\bsx,\pi)}
\ee
for any fixed permutation $\pi$.
Both sides factorise according to the cycles of $\pi$.
The contribution of cycles of length 1,3,4,\dots is identical.
We then obtain an equation for cycles of length 2, namely
\be
\int_{\Lambda^2} \dd x_1 \dd x_2 \e{-\frac1{2\beta} |x_1-x_2|^2 - V(x_1,x_2,x_2,x_1)} = \int_{\Lambda^2} \dd x_1 \dd x_2 \e{-\frac1{2\beta} |x_1-x_2|^2 - \alpha}.
\ee
Using \eqref{V premier ordre}, a few computations give \cite{BU2}
\be
\label{physical alpha}
\alpha = \Bigl( \frac8{\pi\beta} \Bigr)^{1/2} a + O(a^2).
\ee

We emphasise that the approximation consists in retaining only interactions within 2-cycles.
Computations are exact afterwards, at least to lowest order in the scattering length of the interaction potential $U$ between the quantum particles.

\subsection{Pressure and critical density}

We now generalise a bit the model above by considering more general one-body terms, as we have done throughout this article.
Let
\be
H^{(\alpha)}(\bsx,\pi) = \sum_{i=1}^N \xi_\Lambda(x_i-x_{\pi(i)}) + \alpha N_2(\pi),
\ee
with $\xi_\Lambda$ as in Section \ref{secinfcycles}, and $\alpha$ is a positive parameter.

The pressure of this simple model can be computed exactly.
One gets the critical density by analogy with the ideal gas.
We state below a result about infinite cycles for a larger density, see Theorem \ref{thmalpha}.
It remains an open problem to show that infinite cycles are present all the way to the critical density.

For given $\alpha$, the pressure depends on the chemical potential $\mu$ and is given by
\be
\label{une pression !}
p^{(\alpha)}(\mu) = \lim_{V\to\infty} \frac1V \log Z(\Lambda,\mu)
\ee
with $Z(\Lambda,\mu)$ the ``grand-canonical partition function''.
We can define it directly in the Fourier space, by
\be
\label{fpart gc}
Z(\Lambda,\mu) = \sum_{N\geq0} \frac{\e{\mu N}}{N!} \sum_{k_1,\dots,k_N \in \Lambda^*}
\sum_{\pi \in \caS_N} \e{-\alpha N_2(\pi)} \prod_{i=1}^N \e{-\varepsilon(k_i)} \delta_{k_i, k_{\pi(i)}}.
\ee
We need the pressure of the ideal gas
\be
\label{idealpressure}
p^{(0)}(\mu) = -\int_{\bbR^d} \log \bigl( 1 - \e{-(\varepsilon(k)-\mu)} \bigr) \dd k.
\ee

\begin{theorem}
\label{thmpressure}
For $\mu<0$, the limit \eqref{une pression !} exists, and
\[
p^{(\alpha)}(\mu) = p^{(0)}(\mu) - \tfrac12 \e{2\mu} [1-\e{-\alpha}] \int_{\bbR^d} \e{-2\varepsilon(k)} \dd k.
\]
\end{theorem}

Notice that the model is defined only for $\mu<0$, like the ideal gas.
Its derivative at $\mu = 0-$ gives the critical density, and we find
\be
\label{critdensalpha}
\rho_{\rm c}^{(\alpha)} = \frac{\partial p^{(\alpha)}}{\partial \mu} \Big|_{\mu=0-} = \rho_{\rm c}^{(0)} - [1-\e{-\alpha}] \int \e{-2\varepsilon(k)} \dd k.
\ee
The first term of the right side is equal to the critical density of the ideal gas, Eq.\ \eqref{critdens}.
The second term is the correction due to our simple interaction.

The physically relevant situation is $d=3$, $\xi(x) = |x|^2/4\beta$, and $\alpha$ in \eqref{physical alpha}.
We find that, to lowest order in $a$,
\be
\frac{T_{\rm c}^{(a)} - T_{\rm c}^{(0)}}{T_{\rm c}^{(0)}} \approx \tilde c \, \rho^{1/3} a,
\ee
with $\tilde c = 0.37$.
The details of the computations can be found in \cite{BU2}.
This formula can be compared with \eqref{crittempa}.
Our constant has the correct sign, and is about a quarter of the expected constant.
This suggests that the simplified model accounts for some of the effects of interactions on the critical temperature for Bose-Einstein condensation.
Besides, it allows for simple but illuminating heuristics:
Interactions discourage small cycles; then all other cycles are favoured, including infinite cycles.
As a consequence, repulsive interactions increase the critical temperature (or decrease the critical density).

\begin{proof}[Proof of Theorem \ref{thmpressure}]
From \eqref{fpart gc}, we have
\be
Z(\Lambda,\mu) = \sum_{(n_k)_{k\in\Lambda^*}} \prod_{k \in \Lambda^*} \biggl[ \e{-(\varepsilon(k)-\mu) n_k}
\sum_{\pi_k \in \caS_{n_k}} \frac1{n_k!} \e{-\alpha N_2(\pi_k)} \biggr].
\ee
We decomposed the permutation $\pi$ into permutations $(\pi_k)$ for each Fourier mode, and we also used
\be
N_2(\pi) = \sum_{k \in \Lambda^*} N_2(\pi_k).
\ee
Notice that the chemical potential needs to be strictly negative, as in the ideal gas.
We get
\be
\label{palpha}
p^{(\alpha)}(\mu) = \lim_{V\to\infty} \frac1V \sum_{k\in\Lambda^*} \log \biggl[ \sum_{n\geq0} \e{-(\varepsilon(k)-\mu) n}
\sum_{\pi \in \caS_n} \frac1{n!} \e{-\alpha N_2(\pi)} \biggr].
\ee
Let us compute the bracket above.
For given $\pi \in \caS_n$, let $r_j$ denote the number of cycles of length $j$.
Then $\sum_j j r_j = n$, and the number of permutations for given $(r_j)$ is equal to
\[
n! \Big/ \prod_{j\geq1} j^{r_j} r_j!.
\]
The bracket in \eqref{palpha} is then equal to
\[
\begin{split}
&\sum_{n\geq0} \frac1{n!} \sumtwo{r_1,r_2,\dots \geq 0}{\sum_j j r_j = n} \frac{n!}{\prod_{j\geq1} j^{r_j} r_j!} \e{-(\varepsilon(k)-\mu) \sum_j j r_j} \e{-\alpha r_2} \\
= &\sum_{r_1,r_3,r_4,\dots\geq0} \prod_{j=1,3,4,\dots} \frac1{r_j!} \bigl[ \tfrac1j \e{-j(\varepsilon(k)-\mu)} \bigr]^{r_j}
\sum_{r_2\geq0} \frac1{r_2!} \bigr[ \tfrac12 \e{-2(\varepsilon(k)-\mu) - \alpha} \bigr]^{r_2} \\
= &\exp \Bigl\{ \sum_{j=1,3,4,\dots} \tfrac1j \e{-j(\varepsilon(k)-\mu)} + \tfrac12 \e{-2(\varepsilon(k)-\mu) - \alpha} \Bigr\} \\
= &\exp \Bigl\{ -\log (1 - \e{-(\varepsilon(k)-\mu)}) - \tfrac12 \e{-2(\varepsilon(k)-\mu)} [1-\e{-\alpha}] \Bigr\}.
\end{split}
\]
We can insert this into \eqref{palpha}.
In the limit $V\to\infty$ the expression converges to a Riemann integral.
\end{proof}

\subsection{Occurrence of infinite cycles}

Given $\alpha \in [0,\infty]$, the expectation of a random variable on permutations is given by
\be
E_{\Lambda,N}(\theta) = \frac1{Z(\Lambda,N) N!} \sum_{\pi\in\caS_N} \theta(\pi) \e{-\alpha N_2(\pi)}
\int_{\Lambda^N} \dd x_1 \dots \dd x_N \prod_{i=1}^N \e{-\xi_\Lambda(x_i - x_{\pi(i)})}.
\ee
The normalisation $Z(\Lambda,N)$ depends on $\alpha$, although the notation does not make it explicit.
We expect that the claims of Theorem \ref{thminfinitecycles} extend to $\alpha\neq0$, with the critical density given by \eqref{critdensalpha} instead of \eqref{critdens}.
But we only state and prove a weaker claim.

\begin{theorem}
\label{thmalpha}
For all $0<b<1$,
\[
\lim_{V\to\infty} E_{\Lambda,\rho V} (\bsvarrho_{V^b,\rho V}) \geq \rho - \frac4{(1 + \e{-\alpha})^2} \rho_{\rm c}^{(0)}.
\]
\end{theorem}

Of course, the theorem is useful only if the right side is strictly positive.
The proof is similar to Theorem \ref{thminfinitecycles}, but there is one important difference.
We cannot invoke a set of typical occupation numbers, such as $A_\eta$ in Eq.\ \eqref{defAeta}.
We prove below (Proposition \ref{prop4}) that the zero Fourier mode is macroscopically occupied if the density is large enough.
But we need it to be strictly bigger than $\rho_{\rm c}^{(0)}$, itself bigger than $\rho_{\rm c}^{(\alpha)}$.

Let us define
\be
\label{defhn}
h_n(\alpha) = \frac1{n!} \sum_{\pi\in\caS_n} \e{-\alpha N_2(\pi)}.
\ee
Notice that $h_n(0)=1$.
We introduce a probability on $(\Lambda^*)^N \times \caS_N$ that generalises Eq.\ \eqref{probF1}:
\be
P_{\Lambda,N}(\bsk \times \pi) = \frac1{Z'(\Lambda,N) N!} \e{-\alpha N_2(\pi)} \e{-\sum_{i=1}^N \varepsilon(k_i)}
\ee
if $k_i = k_{\pi(i)}$ for all $i$, it is zero otherwise.
Summing over permutations, and over vectors $\bsk$ that are compatibles with $\bsn$, we get
\be
P_{\Lambda,N}(\bsn) = \frac1{Z'(\Lambda,N)} \prod_{k \in \Lambda^*} \e{-\varepsilon(k)} h_{n_k}(\alpha).
\ee

We generalise Lemma \ref{lemkeyFourierproperty}.

\begin{lemma}
\label{lemma1}
We have $Z(\Lambda,N) = C^N Z'(\Lambda,N)$ with $C = \int \e{-\xi}$; and for all $\theta \in \caR_N$,
\[
E_{\Lambda,N}(\theta) = \sum_{\bsn \in \caN_{\Lambda,N}} P_{\Lambda,N}(\bsn) \sum_{\pi \in \caS_N} \theta(\pi) P_{\Lambda,N}(\pi | \bsk)
\]
with $\bsk$ any $N$-tuple such that $\bsn(\bsk) = \bsn$.
\end{lemma}

\begin{proof}
We use Corollary \ref{corFourier} to rewrite the expectation $E_{\Lambda,N}$ in the Fourier space:
\be
\begin{split}
E_{\Lambda,N}(\theta) &= \frac1{Z(\Lambda,N) N!} \sum_{\pi\in\caS_N} \theta(\pi) \e{-\alpha N_2(\pi)} \sum_{k_1,\dots,k_N \in \Lambda^*} \prod_{i=1}^N \delta_{k_i, k_{\pi(i)}} C \e{-\varepsilon(k_i)} \\
&= \sum_{\pi\in\caS_N} \theta(\pi) \sum_{\bsk \in (\Lambda^*)^N} P_{\Lambda,N}(\bsk \times \pi) \\
&= \sum_{\bsk \in (\Lambda^*)^N} P_{\Lambda,N}(\bsk) \sum_{\pi\in\caS_N} \theta(\pi) P_{\Lambda,N}(\pi | \bsk).
\end{split}
\ee
One can sum first over $\bsn$ and then over compatible $\bsk$'s.
\end{proof}

Next we gather some information on the functions $h_n(\alpha)$ defined in \eqref{defhn}.

\begin{lemma}
\label{lemma2}
For $\alpha \in [0,\infty]$, let $\delta = \frac12 (1-\e{-\alpha}) \in [0,\frac12]$.
\begin{itemize}
\item[(a)] $h_n(\alpha) = \sum_{j=0}^{\lfloor \frac n2 \rfloor} \frac1{j!} (-\delta)^j$.
\item[(b)] $1-\delta \leq h_n(\alpha) \leq 1$.
\item[(c)] $\e{-\delta} - \delta^{n/2} / \lfloor \frac n2 \rfloor ! \leq h_n(\alpha) \leq \e{-\delta} + \delta^{n/2} / \lfloor \frac n2 \rfloor !$.
\end{itemize}
\end{lemma}

\begin{proof}
Isolating the contribution of the cycle that contains 1, we get the following recursive relation; for $n\geq2$,
\be
h_n(\alpha) = \frac1n \sum_{j=0}^{n-1} h_j(\alpha) - \frac1n (1-\e{-\alpha}) h_{n-2}(\alpha).
\ee
We also have $h_0(\alpha) = h_1(\alpha) = 1$.
Now the formula in (a) can be proved by induction.
(b) is a consequence of the alternating series in (a).
Notice that $h_2(\alpha) = h_3(\alpha) = 1-\delta$.
(c) follows from the expression
\be
h_n(\alpha) = \e{-\delta} - \sum_{j \geq \lfloor \frac n2 \rfloor + 1} \frac{(-\delta)^j}{j!}.
\ee
Recall that alternating series of decreasing terms are bounded by their first term.
\end{proof}

Let us define $N_{a,n}(\pi) = \sum_{i=1}^n \upchi_{[a,n]}(\ell_i(\pi))$. Note that $N_{a,n}(\pi) = V \bsvarrho_{a,n}(\pi)$. 

\begin{lemma}
\label{lemma3}
Suppose $a>2$, and let $\delta$ as in Lemma \ref{lemma2}. For all $m\geq0$,
\[
\sum_{\pi\in\caS_n} N_{a,n}(\pi) \frac{\e{-\alpha N_2(\pi)}}{h_n(\alpha) n!} \geq (n - a - 2m) ( 1 - \delta^m).
\]
\end{lemma}

\begin{proof}
When summing over permutations, all indexes $i$ in the definition of $N_{a,n}(\pi)$ are equivalent, so that
\be
\sum_{\pi\in\caS_n} N_{a,n}(\pi) \frac{\e{-\alpha N_2(\pi)}}{h_n(\alpha) n!} = n \sum_{\pi\in\caS_n} \upchi_{[a,n]}(\ell_1(\pi)) \frac{\e{-\alpha N_2(\pi)}}{h_n(\alpha) n!}.
\ee
Summing over the lengths of the cycle that contains 1, we get
\be
\begin{split}
\sum_{\pi\in\caS_n} N_{a,n}(\pi) \frac{\e{-\alpha N_2(\pi)}}{h_n(\alpha) n!} &= \sum_{j=a}^n n (n-1) \dots (n-j+1) \sum_{\pi \in \caS_{n-j}} \frac{\e{-\alpha N_2(\pi)}}{h_n(\alpha) n!} \\
&= \sum_{j=a}^n \sum_{\pi \in \caS_{n-j}} \frac{\e{-\alpha N_2(\pi)}}{h_{n-j}(\alpha) (n-j)!} \frac{h_{n-j}(\alpha)}{h_n(\alpha)}.
\end{split}
\ee
We get a lower bound by summing up to $n - 2m$.
By Lemma \ref{lemma2} (c), we have for $2 < j < n-2m$
\be
\frac{h_{n-j}(\alpha)}{h_n(\alpha)} \geq \frac{\e{-\delta} - \delta^m/m!}{\e{-\delta} + \delta^{n/2} / \lfloor \frac n2 \rfloor !} \geq 1 - \delta^m.
\ee
\end{proof}

The last result that is needed for the proof of Theorem \ref{thmalpha} is that our interacting model displays Bose-Einstein condensation, in the sense that the zero Fourier mode is macroscopically occupied.

\begin{proposition}
\label{prop4}
The expectation for the occupation of the zero Fourier mode is bounded below by
\[
\lim_{V\to\infty} E_{\Lambda,N} \bigl( \tfrac{n_0}{V} \bigr) \geq \rho - \frac4{(1+\e{-\alpha})^2} \rho_{\rm c}^{(0)}.
\]
\end{proposition}

\begin{proof}
We proceed as in Appendix B of \cite{Uel1}.
We have
\be
E_{\Lambda,N} \bigl( \tfrac{n_0}{V} \bigr) = \rho - \frac1V \sum_{k \in \Lambda^* \setminus \{0\}} E_{\Lambda,N}(n_k),
\ee
and
\be
\begin{split}
E_{\Lambda,N}(n_k) &= \sum_{j\geq1} P_{\Lambda,N}(n_k \geq j) \\
&= \sum_{j\geq1} \sum_{\bsn \in \caN_{\Lambda,N-j}} \e{-\varepsilon(k) j} \frac{\prod_{k' \in \Lambda^*} \e{-\varepsilon(k') n_{k'}} h_{n_{k'}}(\alpha)}{Z'(\Lambda,N)} \, \frac{h_{n_k+j}(\alpha)}{h_{n_k}(\alpha)}.
\end{split}
\ee
The latter ratio is smaller than $(1-\delta)^{-1}$ by Lemma \ref{lemma2} (b).
By restricting occupation numbers to $n_0 \geq j$, we also have
\be
\begin{split}
Z'(\Lambda,N) &\geq \sum_{\bsn \in \caN_{\Lambda,N-j}} \prod_{k \in \Lambda^*} \e{-\varepsilon(k) n_k} h_{n_k}(\alpha) \frac{h_{n_0+j}(\alpha)}{h_{n_0}(\alpha)} \\
&\geq Z'(\Lambda,N-j) (1-\delta).
\end{split}
\ee
Then
\be
E_{\Lambda,N}(n_k) \leq (1-\delta)^{-2} \sum_{j\geq1} \e{-\varepsilon(k) j} = (1-\delta)^{-2} \frac1{\e{\varepsilon(k)} - 1}.
\ee
It follows that
\be
E_{\Lambda,N} \bigl( \tfrac{n_0}{V} \bigr) \geq \rho - (1-\delta)^{-2} \frac1V \sum_{k \in \Lambda^* \setminus \{0\}} \frac1{\e{\varepsilon(k)} - 1}.
\ee
We get the proposition by letting $V \to \infty$.
\end{proof}

\begin{proof}[Proof of Theorem \ref{thmalpha}]
From Lemma \ref{lemma1}, we have
\be
E_{\Lambda,N}(\bsvarrho_{m,n}) = \sum_{\bsn \in \caN_{\Lambda,N}} P_{\Lambda,N}(\bsn) \sumtwo{\pi\in\caS_N}{k_i = k_{\pi(i)} \, \forall i} \bsvarrho_{m,n}(\pi) \frac{\e{-\alpha N_2(\pi)}}{\prod_{k \in \Lambda^*} h_{n_k}(\alpha) n_k!}.
\ee
Compatible permutations factorise according to Fourier modes, i.e.\ $\pi = (\pi_k)$ with $\pi_k \in \caS_{n_k}$.
Also, $N_2(\pi) = \sum_k N_2(\pi_k)$.
Then
\be
\begin{split}
E_{\Lambda,N}(\bsvarrho_{m,n}) &= \sum_{\bsn \in \caN_{\Lambda,N}} P_{\Lambda,N}(\bsn) \biggl( \prod_{k \in \Lambda^*} \sum_{\pi_k \in \caS_{n_k}} \frac{\e{-\alpha N_2(\pi_k)}}{h_{n_k}(\alpha) n_k!} \biggr) \bsvarrho_{m,n} \bigl( (\pi_k) \bigr) \\
&= \sum_{\bsn \in \caN_{\Lambda,N}} P_{\Lambda,N}(\bsn) \sum_{k\in\Lambda^*} \sum_{\pi_k \in \caS_{n_k}} \bsvarrho_{m,n}(\pi_k) \frac{\e{-\alpha N_2(\pi_k)}}{h_{n_k}(\alpha) n_k!}.
\end{split}
\ee
We keep only the term $k=0$.
Using Lemma \ref{lemma3}, we obtain the lower bound
\be
\begin{split}
E_{\Lambda,N}(\bsvarrho_{V^b,N}) &\geq \sum_{\bsn \in \caN_{\Lambda,N}} P_{\Lambda,N}(\bsn) \frac{n_0 - 3V^b}{V} (1-\delta^{V^b}) \\
&\geq E_{\Lambda,N} \bigl( \tfrac{n_0}{V} \bigr) - 4V^{b-1}.
\end{split}
\ee
The claim follows from Proposition \ref{prop4}.
\end{proof}

\appendix

\section{Macroscopic occupation of the zero Fourier mode.}

In this appendix we investigate the random variable $\bsn \mapsto n_0$ under the 
measure (\ref{probF3}) in the thermodynamic limit $V \to \infty$, $N = \rho V$, for all density parameters $\rho$. We want to show that $n_0 / V$ approaches a limit for each density $\rho$. We will actually show much more, by giving the limiting moment generating function of $n_0 / V$. 
We partly follow Buffet and Pul\'e \cite{BP}, who considered the ideal Bose gas in arbitrary domains.

\begin{theorem}
\label{thmBP}
Let $\rho_0 = \max(0, \rho-\rho_{\rm c})$, with $\rho_{\rm c}$ the critical density defined in \eqref{critdens}.
Then
\[
\lim_{V\to\infty} E_{\Lambda,\rho V}(\e{\lambda n_0 /V}) = \e{\lambda \rho_0}
\]
for all $\lambda \geq 0$.
\end{theorem}

Our first proof applies only when $\rho_c$ is finite.
We give below an argument that completes the proof.

\begin{proof}[Proof when $\rho_{\rm c} < \infty$]
It is shown in \cite{Uel1}, Appendix B, that for all $k \in \Lambda^\ast$, we have 
\be
\label{nicerelation}
P_{\Lambda,N}(n_k \geq j) = \e{-\varepsilon(k) j} \frac{Z'(\Lambda,N-j)}{Z'(\Lambda,N)}.
\ee
Since $P(n_k = j) = P(n_k \geq j) - P(n_k \geq j+1)$, we find for $b > 0$ 
\begin{eqnarray}
E_{\Lambda,N}\left(\e{\nu n_0 }\right) &=& \frac{1}{Z'(\Lambda,N)} \sum_{j=0}^{N} 
\e{\nu j} (Z'(\Lambda,N-j) - Z'(\Lambda,N-j-1)) \nonumber\\
 &=& \frac{\e{\nu N}}{Z'(\Lambda,N)} 
\sum_{j=0}^{N} \e{-\nu j} (Z'(\Lambda,j) - Z'(\Lambda,j-1)).
\label{e lambda n}
\end{eqnarray}
Here, we used the convention $Z(\Lambda,-1) := 0$ and the fact that $P_{\Lambda,N}(n_k \geq N+1) = 0$.
Putting in $N = \floor{\rho V}$ and setting $\nu = \lambda / V$ we obtain
\be
\label{limit candidate}
E_{\Lambda,\rho V} \left(\e{\lambda n_0 / V }\right) = 
\frac{\e{\lambda \rho}}{Z'(\Lambda,\rho V)} 
\sum_{j=0}^{\rho \Lambda} \e{-\frac{\lambda}{V} j} (Z'(\Lambda,j) - Z'(\Lambda,j-1)).
\ee
Above, we wrote $\rho V$ instead of $\floor{\rho V}$ and we will continue to do so, to simplify notation. Now comes the clever insight of Buffet and Pul\'e \cite{BP}: In (\ref{limit candidate}), both $Z'(\Lambda, \rho V)$ and the sum over $j$ can be written as integrals with respect to a purely atomic, $V$-dependent measure $\mu_\Lambda$ on $\bbR^+$; since the functions that are being integrated will not depend on $V$, we only need to study the limit of $\mu_\Lambda$. The measure $\mu_\Lambda$ is given by   
\[
\mu_\Lambda := C_\Lambda \sum_{j=0}^\infty (Z'(\Lambda,j) - Z'(\Lambda,j-1)) \delta_{j/V}
\]
on $\bbR^+$. $\delta_x$ denotes the Dirac measure at $x$, and the constant $C_\Lambda$ will be fixed later in order to obtain a limit measure. From (\ref{limit candidate}) it is now immediate that 
\be \label{measure representation}
E_{\Lambda,\rho V} \left(\e{\lambda n_0 / V }\right) = \e{\lambda \rho} \frac{\int 1_{[0,\rho]}(x) \e{-\lambda x} \mu_\Lambda(\dd x)}{\int 1_{[0,\rho]}(x)  \mu_\Lambda(\dd x)}.
\ee
What makes the idea work is that we can actually calculate the Laplace transform of $\mu_\Lambda$ and take the limit. We have 
\begin{eqnarray*}
 \int_0^\infty \e{-\lambda x} \mu_\Lambda(\dd x) &=& C_\Lambda \sum_{j=0}^\infty \e{-\frac{\lambda j}{V}} 
(Z'(\Lambda,j) - Z'(\Lambda,j-1)) = \\
&=& C_\Lambda (1 - \e{-\frac{\lambda}{V}}) \sum_{j=0}^\infty \e{-\frac{\lambda j}{V}} Z'(\Lambda,j) = \\
&=& C_\Lambda (1 - \e{-\frac{\lambda}{V}}) \exp \left( - \sum_{k \in \Lambda^\ast} \log \left(
1 - \e{-\frac{\lambda}{V} - \eps(k)}\right) \right) = \\
&=& C_\Lambda \exp \left( - \sum_{k \in \Lambda^\ast \setminus \{0\}} \log \left( 
1 - \e{-\frac{\lambda}{V} - \eps(k)}\right) \right).
\end{eqnarray*}
The second equality above is just an index shift, the third is the well-known formula for the pressure of the ideal Bose gas, Eq.\ \eqref{idealpressure}, and the last line follows from $\eps(0) = 0$. The sum in the exponent of the last line is actually quite manageable. By the fundamental theorem of calculus we have for each $k \neq 0$
\[
\log \left(1 - \e{-\frac{\lambda}{V} - \eps(k)}\right) = \frac{1}{V}\int_0^\lambda 
\frac{1}{\e{y/V + \eps(k)} - 1} \dd y + \log \left(1 - \e{- \eps(k)}\right),
\]
and summation over $k$ gives 
\be
\label{div + conv}
\sum_{k \in \Lambda^\ast \setminus \{0\}} \log \left( 
1 - \e{-\frac{\lambda}{V} - \eps(k)}\right)  = 
\sum_{k \in \Lambda^\ast \setminus \{0\}} \log \left( 
1 - \e{- \eps(k)}\right) + \lambda \rho_{\rm c,\Lambda}
\ee
with 
\[
\rho_{\rm c,\Lambda} = \frac{1}{\lambda} \int_0^\lambda 
 \frac{1}{V} \sum_{k \in \Lambda^\ast \setminus \{0\}}
\frac{1}{\e{y/V + \eps(k)} - 1} \dd y
\]
The first term in (\ref{div + conv}) diverges as $V \to \infty$ and defines $C_\Lambda$. The second term converges to the critical density $\rho_{\rm c}$: the integrand is decreasing as a function 
of $y$ and converges to $\rho_{\rm c}$ as a Riemann sum for each fixed $y$, since 
\[
\e{-y/V} \frac{1}{V} \sum_{k \in \Lambda^\ast \setminus \{0\}}
\frac{1}{\e{\eps(k)} - 1} \leq \frac{1}{V} \sum_{k \in \Lambda^\ast \setminus \{0\}}
\frac{1}{\e{y/V + \eps(k)} - 1} \leq \frac{1}{V} \sum_{k \in \Lambda^\ast \setminus \{0\}}
\frac{1}{\e{ \eps(k)} - 1}.
\]
Dominated convergence in $y$ now proves convergence to $\rho_{\rm c}$. We have thus shown that for all $\lambda > 0$
\[
\lim_{V \to \infty} \int_0^\infty e^{- \lambda x} \mu_\Lambda(\dd x) = \e{-\lambda \rho_{\rm c}},
\]
and thus by the general theory of Laplace transforms $\mu_\Lambda \to \delta_{\rho_{\rm c}}$ weakly. 
When used in (\ref{measure representation}), this shows the claim for $\rho > \rho_{\rm c}$. For 
$\rho < \rho_{\rm c}$, both denominator and numerator go to zero, and we need a different argument: we note that by what we have just proved 
\[
\lim_{\rho \searrow \rho_{\rm c}} \lim_{V\to\infty} E_{\Lambda,\rho V}(\e{\lambda n_0 /V})  = 1.
\]
Since the expectation above can never be less than one, all we need to show is monotonicity in $\rho$,
i.e.\ 
\be \label{rho monotonicity}
P_{\Lambda,N+1}(n_0 \geq j) \geq P_{\Lambda,N}(n_0 \geq j).
\ee
For this we use (\ref{nicerelation}) and we obtain
\[
P_{\Lambda,N+1}(n_k \geq j) = P_{\Lambda,N}(n_k \geq j) \frac{P_{\Lambda,N+1}(n_k \geq 1)}
{P_{\Lambda,N-j+1}(n_k \geq 1)},
\]
so it will be enough to show (\ref{rho monotonicity}) for $j=1$. By (\ref{nicerelation}) this means we have to show 
\[
Z'(\Lambda,N)^2 \geq Z'(\Lambda,N-1) Z'(\Lambda,N+1).
\]
Davies \cite{Dav} showed that the finite volume free energy is convex, which proves the inequality above.
\end{proof}

\begin{proof}[Proof of Theorem \ref{thmBP} when $\rho_{\rm c} = \infty$]
We get from Eq.\ \ref{e lambda n}, after some rearrangements of the terms
\be
\label{Hello, Volker!}
\Bigl| E_{\Lambda,N}(\e{\lambda n_0 / V}) - \e{-\lambda / V} \Bigr| = (1 - \e{-\lambda / V}) \e{\lambda\rho} \sum_{i=0}^N \e{-\lambda i / V} \e{-V [ q_\Lambda(i/V) - q_\Lambda(N/V)]}.
\ee
$q_\Lambda(\rho)$ is convex and its limit $q(\rho)$ is strictly decreasing for $\rho < \rho_{\rm c}$ (and $\rho_{\rm c}=\infty$ here).
Beside, we have $q_\Lambda(i/V) - q_\Lambda(N/V) \geq b > 0$, for all $0 \leq i \leq N/2$, uniformly in $V$ and $N = \rho V$.
The right side of \eqref{Hello, Volker!} is then less than
\[
(1 - \e{-\lambda / V}) \e{\lambda\rho} \Bigl[ \sum_{i=0}^{N/2} \e{-Vb} + \sum_{i=N/2}^N \e{-\lambda i / V} \Bigr].
\]
This clearly vanishes in the limit $V\to\infty$.
\end{proof}

Recall the definition of the typical set of occupation numbers $A_\eta$ in \eqref{defAeta}.

\begin{proposition}
\label{proptypoccnum}
For any density $\rho$, and any $\eta>0$,
\[
\lim_{V\to\infty} P_{\Lambda, \rho V}(A_\eta) = 1.
\]
\end{proposition}

\begin{proof}
Let us introduce the following sets of unlikely occupation numbers:
\be
\begin{split}
&A^{(1)} = \bigl\{ (n_k) : \bigl| n_0/V - \rho_0 \bigr| > \eta \bigr\}, \\
&A^{(2)} = \Bigl\{ (n_k) : \sum_{0< |k| < V^{-\eta}} n_k \geq \eta V \Bigr\}, \\
&A^{(3)} = \bigl\{ (n_k) : n_k \geq V^{3\eta} \text{ for some } |k| \geq V^{-\eta} \bigr\}.
\end{split}
\ee
Then
\be
A_\eta^\compl = A^{(1)} \cup A^{(2)} \cup A^{(3)}.
\ee

It follows from Theorem \ref{thmBP} that $P_{\Lambda,\rho V}(A^{(1)})$ vanishes in the limit $V\to\infty$.
Equation \eqref{nicerelation} can be written using free energies as
\be
\label{probnk}
P_{\Lambda,N}(n_k \geq i) = \e{-\varepsilon(k) i} \e{-V (q_\Lambda(\frac{N-i}{V}) - q_\Lambda(\frac N{V}))}.
\ee
We have already mentioned that $q_\Lambda$ is convex, and the limit $q$ is decreasing. Then
\be
q_\Lambda(\rho-\epsilon) - q_\Lambda(\rho) \geq \nu,
\ee
with $\nu \geq 0$. In addition, $\nu>0$ below the critical density, or if the critical density is infinite.
Since
\be
E_{\Lambda,N}(n_k) = \sum_{i\geq0} P_{\Lambda,N}(n_k \geq i),
\ee
we get a bound for the expectation of occupation numbers for $k\neq0$, namely
\be
E_{\Lambda,N}(n_k) \leq \frac1{\e{\varepsilon(k)+\nu} - 1}.
\ee
From Markov's inequality and the bound above, we get
\be
\begin{split}
P_{\Lambda,N}(A^{(2)}) &\leq \frac{\sum_{0< |k| < V^{-\eta}} E_{\Lambda,N}(n_k)}{\eta V} \\
&\leq \eta^{-1} V^{-1} \sum_{0 < |k| < V^{-\eta}} \frac1{\e{\varepsilon(k)+\nu} - 1}.
\end{split}
\ee
The right side vanishes as $V\to\infty$: Indeed, this holds if the critical density is finite even if $\nu=0$;
and we know that $\nu>0$ if the critical density is infinite.

Finally, \eqref{probnk} implies that $P_{\Lambda,N}(n_k \geq i) \leq \e{-\varepsilon(k) i}$.
Since $\varepsilon(k) > a|k|^2$ for small $k$, we have for large $V$
\be
\begin{split}
P_{\Lambda,N}(A^{(3)}) &\leq \sum_{|k| \geq V^{-\eta}} P_{\Lambda,N}(n_k \geq V^{3\eta}) \\
&\leq \sum_{|k| \geq V^{-\eta}} \e{-\frac12 a V^\eta - \varepsilon(k)} \\
&\leq V \e{-\frac12 a V^\eta} \sum_{k \in \Lambda^*} \frac1{V} \e{-\varepsilon(k)}.
\end{split}
\ee
The prefactor of the last line goes to 0, while the sum converges to a finite Riemann integral.
The whole expression vanishes in the limit.
\end{proof}

\section{Convexity and Fourier positivity}

The case $d=1$ of the following lemma was used to provide an example of a one-dimensional system with finite critical density.
The result is certainly known --- we were told that it may go back to P\'olya --- but it is easier to find a proof than a reference.

\begin{lemma}
\label{lemposFourier}
Let $g : (0,\infty) \to (0,\infty)$ such that $\int_0^\infty r^{d-1} g(r) \dd r < \infty$, and such that $r^{d-1} g(r)$ is convex.
Then the function $f(x) = g(|x|)$ on $\bbR^d$ has positive Fourier transform.
\end{lemma}

\begin{proof}
First, we show that for any convex function $h$ on $(0,\infty)$, we have
\be
\label{specialpositivity}
\int_n^{n+1} h(u) \cos(2\pi u) \dd u \geq 0
\ee
for any integer $n \geq 0$.
It is enough to consider the case $n=0$; the general case follows from a change of variables (translates of convex functions are convex).
We have
\be
\label{specposexpanded}
\int_0^1 h(u) \cos(2\pi u) \dd u = \int_0^{1/4} \bigl[ h(u) - h(\tfrac12 - u) - h(\tfrac12 + u) + h(1-u) \bigr] \cos(2\pi u) \dd u.
\ee
We used the fact that
\be
\cos(2\pi u) = -\cos(2\pi (\tfrac12 - u)) = -\cos(2\pi (\tfrac12 + u)) = \cos(2\pi (1-u)).
\ee
We now show that the bracket in \eqref{specposexpanded} is positive.
Let $\alpha = \frac{1/2}{1-2u}$.
Because $h$ is convex,
\[
\begin{split}
h(\tfrac12-u) &= h(\alpha u + (1-\alpha) (1-u)) \leq \alpha h(u) + (1-\alpha) h(1-u), \\
h(\tfrac12+u) &= h((1-\alpha) u + \alpha (1-u)) \leq (1-\alpha) h(u) + \alpha h(1-u).
\end{split}
\]
This proves positivity of the bracket of \eqref{specposexpanded}, hence of \eqref{specposexpanded} and \eqref{specialpositivity}.
The case $d=1$ of Lemma \ref{lemposFourier} follows, since $\widehat f(k) = 2\int_0^\infty f(r) \cos(2\pi |k| r) \dd r$.

For $d\geq2$, let $\theta$ denote the angle between $x$ and $k$, and let $\Xi(\theta) \geq 0$ denote the measure of all remaining angles
($\Xi(\theta) = 2$ for $d=2$, $\Xi(\theta) = 2\pi\sin\theta$ for $d=3$).
Then
\be
\begin{split}
\widehat f(k) &= \int_0^\infty r^{d-1} \dd r \int_0^\pi \Xi(\theta) \dd\theta \, g(r) \cos(2\pi |k| r \cos\theta) \\
&= 2\int_0^{\pi/2} \frac{\Xi(\theta) \dd\theta}{(|k| \cos\theta)^d} \int_0^\infty \dd u \, u^{d-1} g \Bigl( \frac u{|k| \cos\theta} \Bigr) \cos(2\pi u).
\end{split}
\ee
Now $u^{d-1} g(\frac u{|k| \cos\theta})$ is convex in $u$ for given $|k|$ and given $0<\theta<\frac\pi2$ (scaled convex functions are convex).
The latter integral is positive by \eqref{specialpositivity}.
\end{proof}

\end{document}